\documentclass[letterpaper,twocolumn,10pt]{article}
\usepackage{usenix-2020-09}

\usepackage{tikz}
\usepackage{amsmath}

\usepackage{verbatim}
\usepackage{subcaption}
\usepackage{graphicx}
\usepackage{tcolorbox}
\usepackage{color, colortbl}

\usepackage{url}

\usepackage{breakurl}

\newlength{\smallfigwidth}
\newlength{\quartergwidth}
\begin{document}
\date{}

%\title{StormCloud: Testing Variability of Communication in the Cloud}
\title{Cloudy Forecast: \\ How Predictable is Communication Latency in the Cloud?}

%\author{
%{\rm Anonymous Author(s)}
%}

\author{
{\rm Owen Hilyard}\\
University of New Hampshire
\and
{\rm Bocheng Cui}\\
University of New Hampshire
\and
{\rm Marielle Webster}\\
University of New Hampshire
\and
{\rm Abishek Bangalore Muralikrishna}\\
University of New Hampshire
\and
{\rm Aleksey Charapko}\\
University of New Hampshire
}
% end author

\renewcommand{\sectionautorefname}{\S}
\renewcommand{\subsectionautorefname}{\S}
\renewcommand{\subsubsectionautorefname}{\S}

\maketitle

% Disable header and footer on the from page
\thispagestyle{empty}

%------------------------------------
% Absrtact
%------------------------------------

\begin{abstract}
Many systems and services rely on timing assumptions for performance and availability to perform critical aspects of their operation, such as various timeouts for failure detectors or optimizations to concurrency control mechanisms. Many such assumptions rely on the ability of different components to communicate on time---a delay in communication may trigger the failure detector or cause the system to enter a less-optimized execution mode. Unfortunately, these timing assumptions are often set with little regard to actual communication guarantees of the underlying infrastructure -- in particular, the variability of communication delays between processes in different nodes/servers. The higher communication variability holds especially true for systems deployed in the public cloud since the cloud is a utility shared by many users and organizations, making it prone to higher performance variance due to noisy neighbor syndrome. In this work, we present Cloud Latency Tester (CLT), a simple tool that can help measure the variability of communication delays between nodes to help engineers set proper values for their timing assumptions. We also provide our observational analysis of running CLT in three major cloud providers and share the lessons we learned.
\end{abstract}

%-----------------------------------
% Sections
%-----------------------------------

%Introduction
\section{Introduction}
\label{sec:intro}
The reliability and robustness of cloud systems and services are critical to businesses and consumers. Recently, several high-profile cloud outages impacted various aspects of daily life. For instance, a December 2021 AWS outage impacted Amazon delivery service, Roomba smart vacuum cleaners, and streaming services~\cite{aws_dec21_summary,aws_dec21_nytimes}. The financial impact of the majority of such outages far exceeds \$100,000~\cite{outage_cost}.

The software systems deployed in the cloud are designed to be fault-tolerant to prevent smaller failures and outages from getting bigger and impacting users. A typical fault-tolerance mechanism relies on redundancy -- systems store many copies of data and have spare compute capacity for processing that data. However, systems also rely on timing assumptions to use this redundancy. For example, a failure of some component needs to be detected before a system may reconfigure to switch to a redundant copy. Such detection usually relies on some timeout or lease mechanism, configured with some timing assumption about expected message delivery. Timing assumptions also play an important role in performance-related optimizations of many protocols and algorithms~\cite{spanner,epaxos_revisited,accord,rabia,copilots}. Whenever the timing assumption holds, the system may operate in a faster mode. Violating the assumption may cause the system to degrade into a less optimized mode. As a result, failures of timing assumptions present a significant obstacle to reliability, as systems may have trouble detecting failures, experience degraded performance/liveness, or both. 

Recently, many algorithms started to rely on increasingly tighter timing assumptions for message delivery. For instance, Copilots~\cite{copilots} relies on 1-millisecond and 10-millisecond timeouts to perform reconfiguration actions. The protocol needs such timeouts to detect slow machines and reconfigure to make progress despite the slow node, ensuring a more predictable performance in case of slowdowns and gray failures~\cite{gray_failures,fail_slow_hotos,fail_slow_hw}. Many other algorithms and systems rely on tight timing assumptions for better concurrency control. The idea for such time-augmented concurrency control is to delay some operation(s) to either ensure safety~\cite{spanner} in the presence of potentially conflicting/concurrent operations or to avoid doing a more expensive conflict resolution~\cite{rabia,epaxos_revisited,accord}. For instance, EPaxos Revisited~\cite{epaxos_revisited} state machine replication protocol expects that nodes see most messages for a replication round by some deadline; if communication takes longer, the protocol may experience a conflict situation when it is uncertain about the order of replicated commands, causing an expensive conflict resolution procedure to kick in. Naturally, these systems cannot afford to wait for longer than strictly necessary and must resort to the tightest possible assumptions on message delivery latency. 

Cloud is a shared environment, with many tenants running their systems and applications in isolated virtualized environments. This isolation often makes it look like each tenant operates with their dedicated resources provided by the cloud provider. However, in reality, tenants share the physical resources of the cloud provider, ranging from servers to networking, storage, and more. Such sharing leads to noisy neighbor syndrome ~\cite{MS_noisy_neighbor,noisy_neighbor_detection,fail_fast_is_failing_fast}, a situation when one tenant may monopolize some shared resource, even if for a very brief time, starving other tenants and negatively impacting their performance. Modern cloud manages the problem decently well at larger time frames -- tenants' ability to burst above their nominal resource allocation is often constrained in time, while various schedulers and cluster management systems~\cite{protean,managing_overloaded_hosts,borg} work hard to achieve a balanced placement of tenants and work within the cluster or datacenter. 

However, with tight expectations on the timing and delivery of messages, even short-lived performance variations of the underlying infrastructure can have a significant negative impact on systems, especially if such variations are frequent. It is also important to note that the timing of message delivery spans more than just the data center network, as several other components impact the latency perceived by the applications and systems, such as the OS kernel, hypervisor, and the applications themselves. However, from the perspective of the application, it matters very little whether the messages get delayed in the network or OS kernel -- the end result is delayed and/or unpredictable message delivery. Thus, the overall cloud's virtualized infrastructure, and not just the cloud network, impacts the application-perceived communication timing. 

To understand the impacts of unpredictable communication timing, consider the example of a 10-millisecond timeout used in Copilots~\cite{copilots} in a cloud environment whose latency spikes above 10 ms once every minute. In such a scenario, the system will likely enter a once-per-minute reconfiguration loop to transition from two co-leaders down to one, paying the associated price in excess resource usage and possibly performance degradation. However, a less tight timeout could have avoided this unnecessary reconfiguration loop. Similarly, systems that rely on message delivery assumptions for concurrency management will go through cycles of high and low performance when communication timing assumptions are violated. Whenever a system uses resources to reconfigure or run in a less-optimized mode, it loses the opportunity to do useful work. These behaviors can take variations in one resource (i.e., constrained communication) and transform and amplify them into variations in another resource (i.e., increased CPU and memory usage to perform reconfiguration or go through a non-optimized execution path). This kind of transformation may even create a feedback loop and cause metastable failure~\cite{metastable_osdi,metastable_hotos}. When the system enters a less performant state, it can cause work to queue up, time out, and retry, creating even more work and exacerbating performance problems. 

Unfortunately, publicly available information on communication latencies in the cloud is scarce, impeding the decision-making process regarding critical timing assumptions. There are a plethora of websites and services that provide some information about average or median latency between cloud regions~\cite{websites_ping1, websites_ping2,websites_ping3}, but these services often do not provide region-local latencies or expose finer statistics on latency distributions to help engineers estimate how often, and by how much the communication latency may deviate from the average or median. As a result, most literature pulls these communication latency timing assumptions out of thin air. For example, the aforementioned Copilots~\cite{copilots} work runs on a dedicated cluster with average inter-node communication latency of 0.25ms, making 1 ms timeout a plausible assumption. 

In this work, we study the predictability of communication in the cloud to understand and empirically justify the timing assumptions engineers and designers make when working on cloud-native services and applications. To that extent, our contributions are two-fold. First, we present CLT, a simple open-source tool\footnote{Available on GitHub: \url{https://github.com/UNH-DistSyS/UNH-CLT}.} to collect communication latency data across many cloud VMs. Second, we use CLT to study the communication latency patterns between VMs in three large cloud providers: Amazon Web Services (AWS), Google Compute Platform (GCP), and Microsoft Azure. 

Our tool, Cloud Latency Tester or CLT for short, is a simple echo-like application that can deploy to many VMs in different parts of the cloud, such as different placement groups, availability zones, or regions. The tool runs TCP traffic of configurable payload size and frequency between the VMs and collects the round-trip latency between all VM pairs. CLT also includes tools and scripts to process raw data and extract valuable statistics for desired pairs or sub-clusters of VMs (i.e., creating a latency histogram for all node pairs in the same availability zone).

We used CLT to collect communication latency data in three large cloud providers. In particular, we look at the communication RTT between VMs in different deployment configurations, such as VMs in the same availability zone (AZ) or across AZs and regions. This data gives a lot of insights into the predictability of cloud communication in several common situations. For instance, we observe the potential for significant tail latency applications may experience in the cloud. The 99.99\textsuperscript{th} percentile tail latency for VMs in the same subnet of the same AZ is as much as 36$\times$ higher than the average latency, while maximum RTT reaches as much as 2900$\times$ the average. Considering that the 99.99\textsuperscript{th} percentile is not that rare (roughly every 10,000\textsuperscript{th} round-trip communication exchange), such high-tail latency can significantly impact applications relying on tight timing and communication latency assumptions. We also notice significant latency variations (as much as 7\% change in 10 minutes) throughout the day across all tested clouds, suggesting a substantial impact from other cloud workloads/tenants on latency.

%Background
\section{Background}
\subsection{Latency in Cloud Systems}
\begin{figure}
    \centering
    \includegraphics[width=0.9\linewidth]{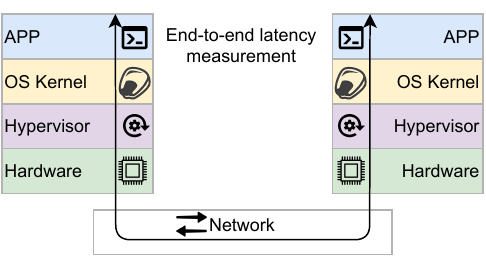}
    \caption{End-to-end latency in cloud systems is impacted by network, hardware, hypervisors, OS kernel, and system itself.}
    \label{fig:end-to-end-latency}
    \vspace{-2mm}
\end{figure}

The end-to-end communication latency in networked applications and systems consists of more than just the network latency between the nodes. Server hardware, virtualization stack, operating system, and the application itself may introduce additional overheads and jitter, as shown in ~\autoref{fig:end-to-end-latency}. Moreover, the relative contribution of these different components may change depending on the system's load and corresponding queuing effects~\cite{harchol2013performance}, operating system, and hypervisor choice. For instance, the Linux Kernel TCP stack can add between 20 to 110,000 microseconds to packet processing compared to kernel-bypass technologies; furthermore, the kernel's TCP stack is significantly more variable, with a latency standard deviation of as much as two orders of magnitude higher than that of kernel bypass solutions \cite{kaufmannTASTCPAcceleration2019, gehbergerPerformanceEvaluationLow2018, jeongMTCPHighlyScalable2014}. Similarly, virtual machine hypervisors can also reduce performance~ \cite{abeniUsingXenKVM2020} across the board, with TCP latency degrading as much as 28\%~\cite{enbergPerformanceEvaluationHypervisor2016}. The compute-intensive performance can also decrease substantially; for example, in~\cite{dingDiagnosingVirtualizationOverhead2015} authors observe as much as 50\% performance penalty for dedup benchmark. Naturally, systems that use more compute capacity can experience longer queuing delays~\cite{harchol2013performance}, impacting the observed end-to-end communication latency. 

\subsection{Noisy Neighbor Syndrome}
Within cloud computing, multiple users or tenants often share the same physical resources partitioned into smaller units such as virtual machines, containers, functions of serverless computing, or even subscriptions to shared services. Ideally, the resources should be distributed fairly among all tenants to ensure smooth and consistent performance---those who pay or subscribe for more resources should be able to get them when needed. However, such fair isolation of tenants may not always work, especially at the smaller time scales, leading to the noisy neighbor syndrome ~\cite{steal_cpu,MS_noisy_neighbor}. The noisy neighbor syndrome refers to the impact some tenants may have on others through the variability of their work by occasionally hogging more resources, such as CPU, disk IOPS or network bandwidth. For example, many services allow bursting---a short-term increase in capacity over some baseline. If too many tenants need to burst at the same time, they may ask to consume more resources than available, impacting each other and potentially other non-bursting tenants. 

Cloud providers have made significant efforts to mitigate the noisy neighbor problem using various approaches, such as resource isolation~\cite{Resource_Management_for_Isolation_Enhanced_Cloud_Services}, fair resource allocation~\cite{fair_resource_allocation,fair_resource_allocation_jp}, and resource scaling out. Moreover, researchers have invested considerable resources in detecting~\cite{noisy_neighbor_detection, Ordozgoiti2017DeepCN, Cloud_Performance_Variability_Prediction} or reducing~\cite{Affinity-Aware_Scheduler, MittOS} the impact of noisy neighbors. Despite the efforts to mitigate the noisy neighbor problem, it may continue to impact the cloud, especially at smaller time scales before the mitigations and isolations can kick in.

%CLT
\section{Cloud Latency Tester}
In this section, we describe Cloud Latency Tester (CLT), a simple tool to measure and study the inter-VM end-to-end communication latency (\autoref{fig:end-to-end-latency}) in the cloud. CLT can deploy on some arbitrary number of VMs in the cloud, covering a desired topology, such as VMs in the same AZ and across different AZs. After deployment, the tool starts a continuous echo-like message exchange in the cluster with messages of configurable size and frequency. CLT records the observed round-trip latencies at each node for further analysis. We implemented CLT in Go~\cite{go}, a popular language for distributed applications and services.

The main goal we want to achieve with CLT is realistic testing of end-to-end communication latency in the cloud environment. This goal partly influenced our language choice and the overall design of the tool. 
Our tool uses a common garbage-collected language that relies on an OS-provided network stack. While a more precise measurement of network latency may be possible by using a non-GC language, such as C++ or Rust, deploying on bare-metal instances to avoid hypervisor, and using kernel-bypass technologies, such as DPDK~\cite{dpdk}, we believe that such comparison would be less representative of an average app or service.

We design CLT to run continuously for prolonged periods. The system operates with many nodes, and each node gets deployed on a separate VM in the cloud. A node has two roles: a sender and a receiver. The sender operates in rounds and broadcasts a message consisting of a round number, some configurable-size payload, and its identity to all the receiver nodes. Upon receiving such a message, the receiver nodes echo back the payload to the sender. The sender ultimately receives the echoes from all receivers and records the round's latency for each receiver. As such, in each round, the sender records multiple latency observations, one for each node in the CLT cluster. Each node, serves both sender and receiver roles, allowing us to record detailed latency observations between all node pairs in the cluster. We use TCP for message exchange as one of the most common communication standards. We use the TCP\_NODELAY option to disable Nagle's algorithm~\cite{nagle,go-turn-off-nagle} and obtain more accurate round-trip latency. 

Each node contains a measurement component to record the round-trip latency for each observation. The challenge here is to isolate the data recording from the workload that runs the message exchange---we do not want the act of recording measurements to impact the communication latency. To that order, our measurement component collects data in memory before periodically flushing it to local storage. Specifically, we flush data to a CSV file in a separate goroutine from recording the new data, such that data collection can work in parallel with flushing already collected data while having little impact on each other, given a sufficient number of available cores in the underlying VM. The tool can also completely bypass flushing data to disk while collecting measurements---each observation needs only 52 bytes of memory, allowing collecting over 20,000,000 latency observations with just 1 GB of memory. In this case, the collected measurements are kept in memory until after the end of the experiment, at which point they get written to local storage. 

CLT also includes helper scripts to deploy, start, stop nodes, and retrieve the data from the VMs in the cluster. Furthermore, we provide rudimentary data-processing capability to convert raw data into histograms and compute statistics, such as average, median, and latency percentiles.

%Eval
\section{Evaluation of Communication in the Cloud}
\label{sec:eval}

% cloud comparison (AWS, GCP, Azure) at low communication rate (100 msg/s) and small (2vCPUs) VM
% - same AZ
%   - different subnet
%   - same subnet
%   - talking to yourself
% - Across AZs
% - Across regions

% Bigger VM (4vCPUs) (at least in the same AZ)
% higher communication rate (300 msg/s) (at least in the same AZ)
% starting VMs at once vs at different time (at least in the same AZ)
% different payload size (in the same AZ and across AZs in the same region): 512 bytes, 1024 bytes, 2048 bytes, 4096 bytes, 8192 bytes. 1 hour each
% different MTU ? 
With the help of CLT, we evaluate the round-trip communication latency and its reliability/predictability in three public clouds: Amazon Web Services (AWS), Google Compute Platform (GCP), and Microsoft Azure. The cloud offers great flexibility in deploying, running, and managing virtualized infrastructure, making it impossible to test all conceivable deployment scenarios. As a result, our goal was to provide as much information relevant to a typical deployment. 

\subsection{Testing Setup}

\begin{figure}
    \centering
    \includegraphics[width=0.9\linewidth]{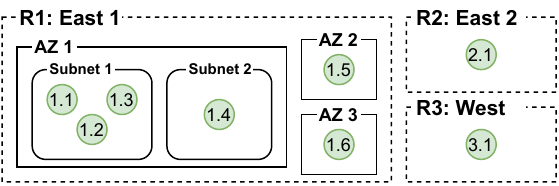}
    \caption{CLT deployment in the cloud.}
    \label{fig:cloud_setup}
    \vspace{-2mm}
\end{figure}

% TODO: consider revising to state our goals first, and then explain how setup achieves these goals
For our experiments, we used a similar setup in each of the three clouds, illustrated in~\autoref{fig:cloud_setup}. In all three clouds, we used a US East region as the main region (East1). In this region, we deployed 6 VMs in 3 different availability zones (AZs), such that AZ1 had 4 VMs in two different subnetworks with cross-subnet communication setups. Two other nodes were in the remaining AZs. This regional setup allowed us to test the communication within the same subnet in the same AZ, across different subnets in the same AZ, and across 3 AZs, representing a few standard deployment configurations~\cite{aurora,dynamodb}. Having 3 nodes in the same AZ and across 3 AZs also enabled us to study the quorum latency, as quorums are typically used in systems to mask the impact of slow or failed nodes. We also deployed two more nodes in two remote regions, referred to as East2 and West. 

We deployed CLT on VMs running under a fresh install of Ubuntu 22.04 LTS, one of the more popular Linux distributions. We compiled the tool with Go 1.20.1. We used a common type of x86-based VMs on each cloud. All VMs ran on comparable Intel CPUs and were configured with 2 vCPUs and 8GB of memory. Specifically, we used m5.large, e2-standard-2, and Standard D2s v3 VMs from AWS, GCP, and Azure respectively. In all cases, dedicated instances were used to avoid any potential issues with spot instances. These VMs are popular choices in the cloud, and for example, D2s v3 appears in a ``most used by Azure users'' section of the Azure portal at the time of our study. The selected VMs also come with a comparable level of networking performance, at least on paper, with AWS offering potentially being the most limited. AWS advertises up to 10 Gbps bandwidth in a burst for its m5.large VM, while providing a base-level bandwidth of only 0.75 Gbps~\cite{aws-m5-specs,aws-m5-bandwidth}, GCP e2-standard-2 VM provides ``maximum egress bandwidth'' of 4 Gbps~\cite{gcp-e2-specs}, and Azure D2s V3 has ``expected network bandwidth'' of 1000 Mbps~\cite{azure_d3_specs}. 

For our base workload, we configured CLT to perform 100 measurement rounds per second at each node in the cluster shown in~\autoref{fig:cloud_setup}. The default payload size was set to 1024 bytes, with 8 nodes in the cluster, resulting in an overall traffic of 1.6 MB/s in each direction (13 Mbps) at each node. These testing parameters also resulted in roughly 15\%-17\% CPU utilization in our VMs, leaving plenty of headroom. We picked this workload to ensure we collect data at sufficient resolution without straining the VMs, as we want to observe the best possible scenario for end-to-end round-trip communication delays. Increasing the load on VMs, for example by collecting more samples each second, can increase queuing effects for the compute and network resources, resulting in higher variability. While real systems are expected to use the most out of their resource, the load characteristics of each system are unique due to the nature of the system and desired SLOs. As such, we want to collect the best-case measurement with the expectation that adding more load to the distributed system can worsen the end-to-end latency. 

\begin{figure*}[t]
  \centering
    \begin{subfigure}{.85\smallfigwidth}
    \includegraphics[width=\linewidth]{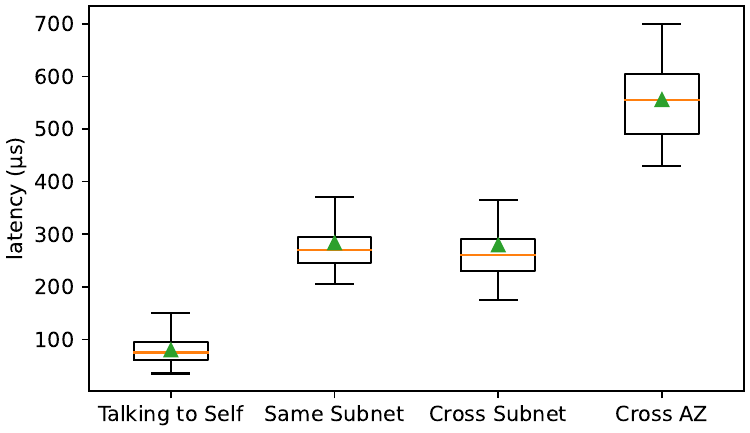}
    \caption{AWS}
    \label{fig:latency_boxplot_aws}
  \end{subfigure}
  \begin{subfigure}{.85\smallfigwidth}
    \includegraphics[width=\linewidth]{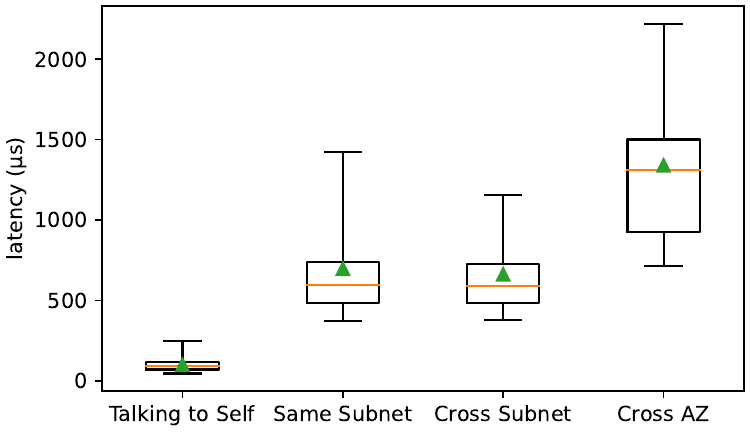}
    \caption{Azure}
    \label{fig:latency_boxplot_azure}
  \end{subfigure}
  \begin{subfigure}{.85\smallfigwidth}
    \includegraphics[width=\linewidth]{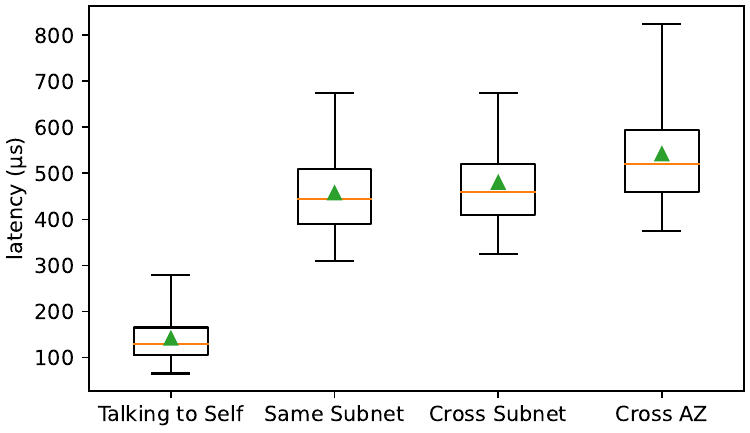}
    \caption{GCP}
    \label{fig:latency_boxplot_gcp}
  \end{subfigure}
    \begin{subfigure}{.5\smallfigwidth}
    \includegraphics[width=\linewidth]{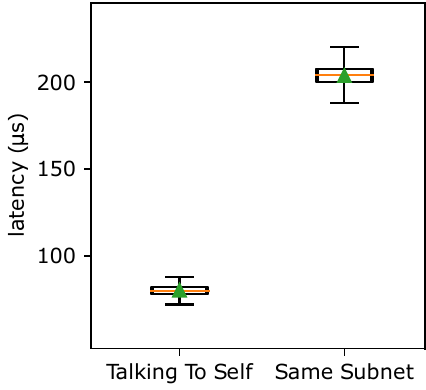}
    \caption{Bare-Metal Rack}
    \label{fig:latency_boxplot_ccl}
  \end{subfigure}
  
  \caption{The box-plot summary for round-trip latency over a 6-hour interval, starting at 2 pm EST on a weekday. The box denotes the IQR, with the middle line designating the median, while the whiskers represent the 5\textsuperscript{th} and 95\textsuperscript{th} percentiles. The green triangle shows the average latency. Note that we are not comparing clouds, so the subfigures have different scales to better show data for each cloud provider. Figure (d) shows latency summary for in-house cluster. }
  \label{fig:latency_agg_boxplot}
  \vspace{-2mm}
\end{figure*}

Since CLT operates in rounds, we can collect point-to-point communication latency between any two nodes in the cluster for each round. We can also collect the quorum latency for each round. Quorums are powerful abstractions to mask slow nodes~\cite{quoracle,naor1998load}, with many systems and services~\cite{spanner,mongodb,cassandra,zookeeper,documentdb,chubby,dynamo} relying on them. In our observation study, we collect data on two quorums -- a majority quorum of nodes in the same AZ (i.e., a majority quorum formed from nodes 1.1, 1.2, and 1.3) and a majority quorum for nodes across 3 AZs (i.e., nodes 1.1, 1.5, and 1.6 as one possible group of such cross-AZ nodes). Unless otherwise stated, most of the data we present comes from a 6-hour run taking place on a weekday between roughly 2 pm and 8 pm in April. % To ensure reproducibility, we have repeated the 6-hour run twice, and also had another 1-hour run, all happening on different weekdays at roughly the same time. Unless otherwise noted, we present the data from the initial 6-hour run.

Public cloud vendors use different hypervisors, hardware, and network topologies inside their data centers. Since we cannot control or often directly observe those components, we treat the entire cloud as a black box. Such a lack of observability makes it impossible to tell which components introduce latency and to what extent. As a result, any explanations for the observed behaviors are, at best, educated guesses that we keep to a minimum. This, however, does not diminish the importance of the observations for latency-sensitive applications and systems since observed latency variations and behaviors may impact their performance.  

\subsection{Threats to Validity}
\label{sec:validity}
Despite our utmost efforts to obtain the most comprehensive, accurate, and reliable data, we still face considerable challenges and limitations that can impact the validity of our observations. While we think that our results can serve as a baseline for discussing the variability of communication in the cloud, we refrain from drawing larger conclusions and generalizations and comparing different cloud providers due to the limited scope of our observations---we used specific VM types during a relatively short period and in a limited number of regions and availability zones. 

\textbf{State of underlying infrastructure:} The underlying infrastructure serves as the foundation for the performance and reliability of cloud service providers' systems. This includes hardware infrastructure, such as networks and servers, and the software infrastructure to support and manage these facilities. The state of the infrastructure may be affected by various factors, such as hardware failures~\cite{Characterizing_cloud_computing_hardware_reliability} and resource contention~\cite{resource_contention}. These issues can impact our observations, limiting our ability to draw definitive conclusions when evaluating cloud services. Furthermore, while we make the best efforts to monitor publicly available outage data, some smaller issues that may impact communication latencies and their variations may not be reported to the public. 

\textbf{Sample size:} Sample size directly affects the significance and reliability of research results. Ideally, we would collect many weeks' worth of observations across all regions, AZs, and VM sizes. However, this is prohibitively expensive and time-consuming. Instead, we focused on providing a sufficiently long baseline and ensuring its reproducibility over multiple runs. While most data we report comes from a single 6-hour run (138 million data points), we conducted multiple such runs and a few smaller ones to ensure reproducibility. Longer continuous runs, however, can differ substantially from our observations due to the diurnal patterns (we capture a hint of such possibility), business patterns, such as the difference between weekdays and weekends, or special events, such as Black Friday. Additionally, our experiments are limited to VMs and do not reflect the performance and reliability of communication in other cloud services, such as serverless compute platforms or container offerings. 

\textbf{Representation of communication patterns:} In cloud systems, communication patterns refer to the ways components interact and collaborate. Communication patterns in production systems and services can be highly complex and varied and may not be properly represented in our simplified observation study. Our study focuses on point-to-point communication between pairs of nodes. We also explore quorum-style communication, in which a sender node expects some threshold of replies before considering the round complete. While these two communication styles represent some basic building blocks, large systems with many components interacting can add further complexity, delays, and variability. 

\textbf{Latency Contributions of Individual Components:} Our observation study is a black-box approach, as such, we have no insight into how much each of the components described in~\autoref{fig:end-to-end-latency} has contributed to the latency or its variability. Arguably, one can look at the self-loop communication to at least normalize for the overheads of virtualization, OS kernel, and the app. However, we believe knowing the contributions of individual components is not necessary for system and application designers/engineers -- the systems deployed in the cloud must account for latency/timing assumptions of their end-to-end infrastructure stack. 

\subsection{Regional Communication}

\begin{figure}[h]
  \centering
  \begin{subfigure}{\linewidth}
    \includegraphics[width=\linewidth]{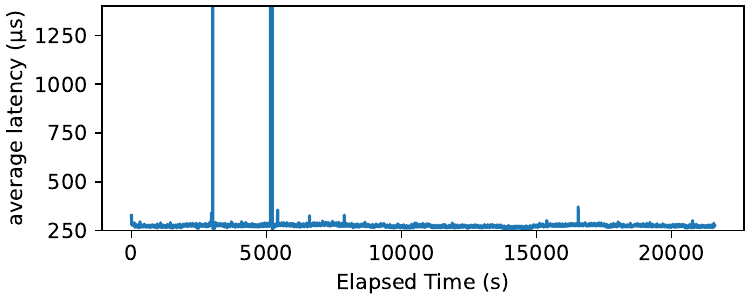}
    \caption{AWS}
    \label{fig:same_subnet_aws}
  \end{subfigure}
  \begin{subfigure}{\linewidth}
    \includegraphics[width=\linewidth]{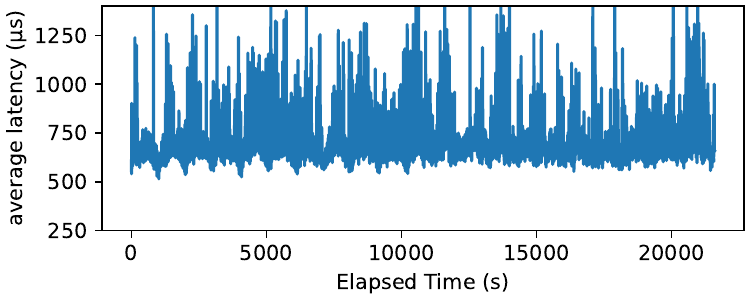}
    \caption{Azure}
    \label{fig:same_subnet_azure}
  \end{subfigure}
  \begin{subfigure}{\linewidth}
    \includegraphics[width=\linewidth]{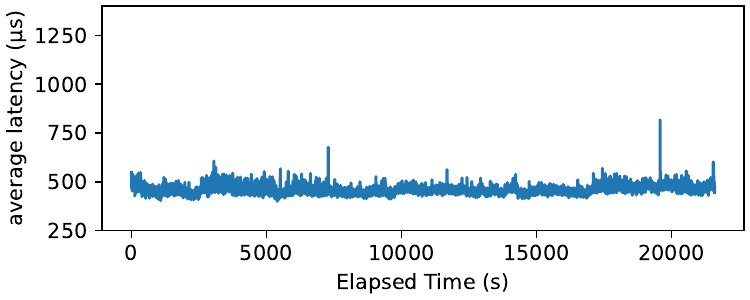}
    \caption{GCP}
    \label{fig:same_subnet_gcp}
  \end{subfigure}
  \begin{subfigure}{\linewidth}
    \includegraphics[width=\linewidth]{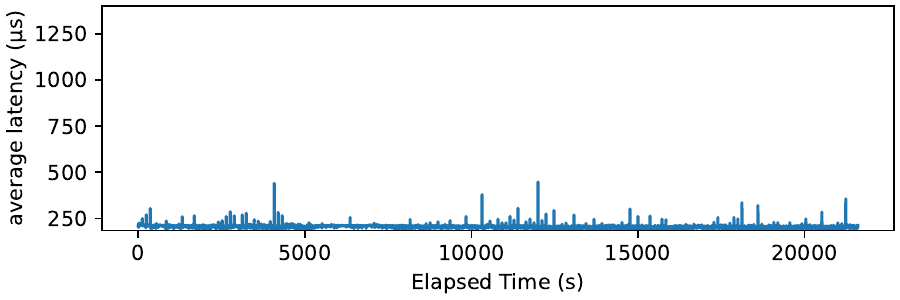}
    \caption{Private Bare-Metal Rack}
    \label{fig:same_subnet_ccl}
  \end{subfigure}
  \caption{Round-trip latency between nodes in the same subnet of the same AZ in East1 region over a 6-hour interval, starting at 2 pm EST on the weekday.}
  \label{fig:latency_same_subnet}
\end{figure}

In the first set of observations, we examine the round-trip communication latency in a regional setup. Referring to~\autoref{fig:cloud_setup}, we look at communication between nodes 1.1, 1.2, and 1.3 for latency within a single availability zone and subnet (referred to as ``Same Subnet'' later in the paper). The latency between node pairs $<$1.1, 1.4$>$, $<$1.2, 1.4$>$, and $<$1.3, 1.4$>$ provide insight into communication delays when two nodes are in different subnets, but still in the same AZ, referred to as ``Cross Subnet''. Finally, node pairs $<$1.1, 1.5$>$, $<$1.1, 1.6$>$, $<$1.2, 1.5$>$, $<$1.2, 1.6$>$, $<$1.3, 1.5$>$, $<$1.3, 1.6$>$, $<$1.4, 1.5$>$, $<$1.4, 1.6$>$ give us communication latency between nodes located in different AZs of the same region.  

%~\autoref{fig:latency_agg_boxplot} is a box plot which summarizes various statistic about the clouds. The green triangle is the median, and the orange line is the mean. The bottom and top of the rectangle that the mean is inside of represent the 25th and 75th percentiles respectively. The error bars represent a 95\% confidence interval. One very interesting insight from this figure is that moving across subnets appears to have a relatively minor impact on latency across all clouds, but moving across AZs can have substantial impacts, which required a different scale for \autoref{fig:cross_az_boxplot}.
We illustrate the summary of our regional experiment in~\autoref{fig:latency_agg_boxplot}. The figure shows the boxplots for four observations in each cloud. ``Talking to Self'' refers to a node using TCP to send a message to itself, ``Same Subnet'' designates the communication within the same subnet of an AZ, while ``Cross Subnet'' shows the communication latency between nodes in different subnets of the same AZ. Finally, ``Cross AZ'' represents the communication between nodes in different AZs of the same region. The figure shows round-trip latency, with the box showing IQR (25th to 75th percentile range) with the median line in between. The whiskers show the 5th and 95th percentiles, and the triangle designates the mean. 

The figure illustrates a handful of important points about the reliability of communication latency in the cloud. One of our initial hunches was that having nodes in different subnets of an AZ may have a noticeable impact on latency, but this is not entirely the case in all clouds. Similarly, we expected a significant difference between communication latency within the same AZ and across AZs. While this is the case, GCP's cross-AZ round-trip communication latency appears very close to its same AZ latency. Note that while the latency distributions have substantial overlap, there is still a statistically significant difference between the ``Same Subnet'' (which has nodes in the same AZ) and ``Cross AZ'' samples (two-sample student's t-test with $t=668.952$ and $p=0$) in GCP. Furthermore, even comparing ``Same Subnet'' with ``Cross Subnet'' shows a significant difference in means for all clouds! However, as the difference between means does not exceed 32 microseconds, it is negligible in practice.

\begin{tcolorbox}[left=1mm,right=1mm,top=1mm,bottom=1mm]
\textbf{Lesson 1:} Nodes in different subnets of the same availability zone are likely to communicate as fast as nodes within the same subnet.
\end{tcolorbox}

Another important observation we make from the overview data (\autoref{fig:latency_agg_boxplot}) is about the latency of self-loop when a node uses its TCP interface to send messages to itself. While sending messages to self is substantially faster than talking to any other node, the time delay is not insubstantial. In fact, at the tail end, all clouds produced observations as high as the latency between nodes in the same subnet/AZ. 

\begin{tcolorbox}[left=1mm,right=1mm,top=1mm,bottom=1mm]
\textbf{Lesson 2:} Talking to self is substantially faster than talking to other nodes, but this is not a guarantee, as in 2 out 3 clouds (GCP and Azure) we observed that 99.99\textsuperscript{th} percentile round-trip latency of talking to self exceeds the mean latency across nodes in the same subnet/AZ. This finding confirms an observation from a few years ago~\cite{taxdc}.
\end{tcolorbox}

Finally, in ~\autoref{fig:latency_boxplot_ccl} we show the same subnet latency summary for a non-cloud environment. This experiment was conducted on our own bare-metal cluster consisting of 16 core AMD EPYC machines connected with a 25Gbps network. CLT did not have access to all cores on each server. Overall, we observe similar results in a talking-to-self experiment, and overall less variance in the same-subnet experiment than most cloud providers.

%% \input{figures/processed_latency_figures/all_figures}

%Figure 1: three vertical panes, one for each cloud, showing birds-eye view (i.e., 15 or 30-second granularity) of the latency in the following pairs: (1.1 <-> 1.2, 1.1 <-> 1.3, 1.2 <-> 1.3.)
\subsubsection{Same AZ Round-Trip Latency}

\autoref{fig:latency_same_subnet} provides a high-level view of the communication latency in the same subnet across a single AZ for all three clouds over an entire 6-hour run starting at roughly 2 pm EST and ending at roughly 8 pm EST. The figure shows the average latency aggregated over 30-second windows. Such aggregation hides some finer latency fluctuations but can capture larger details on the stability of communication latency. This figure excludes the communication latency of nodes talking to themselves and only counts the latency between two distinct nodes in the same subnet of the same AZ. Over this 6-hour interval, we have collected roughly 13 million data points on communication in the same subnet of an AZ.

AWS 30-second averages are relatively stable, as seen in ~\autoref{fig:same_subnet_aws}. However, there are a couple of notable exceptions, as we observe spikes that push the average 30-second window round-trip latency above 1000 microseconds. The spike at roughly 5200 seconds of elapsed time is an especially interesting one. This latency spike is due to one node (1.3) experiencing some networking trouble. This node recorded the same latency of 832.83 ms for all its peers in one communication round, including the nodes in other regions. 
The same node also experienced a three-fold increase in average latency while talking to itself around that time and an elevated latency ($\geq$800 ms) when talking to some peers in other rounds within a few seconds around the big spike, after which the latency returned to normal. This may be an example of some networking or hypervisor problem, largely affecting the ability to receive incoming messages.

\begin{figure*}
  \centering
  \begin{subfigure}{\smallfigwidth}
    \includegraphics[width=\smallfigwidth]{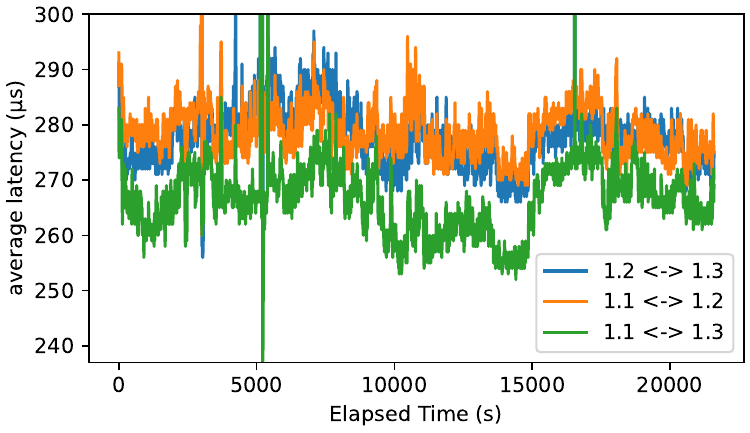}
    \caption{AWS}
    \label{fig:pairwise_latency_same_subnet_aws}
  \end{subfigure}
  \begin{subfigure}{\smallfigwidth}
    \includegraphics[width=\smallfigwidth]{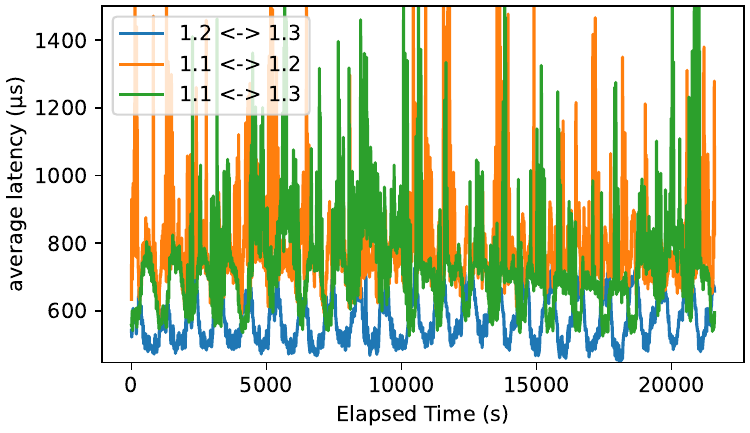}
    \caption{Azure}
    \label{fig:pairwise_latency_same_subnet_azure}
  \end{subfigure}
  \begin{subfigure}{\smallfigwidth}
    \includegraphics[width=\smallfigwidth]{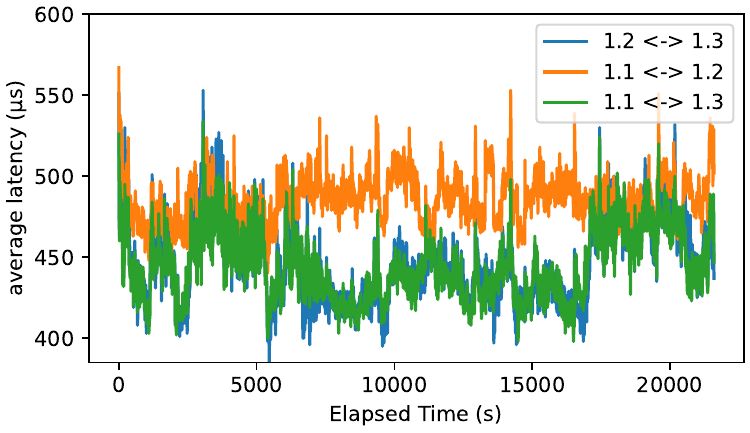}
    \caption{GCP}
    \label{fig:pairwise_latency_same_subnet_gcp}
  \end{subfigure}
  
  \caption{Same Subnet/AZ round-trip latency over a 6-hour interval for individual pairs of nodes.}
  \label{fig:pairwise_latency_same_subnet}
\end{figure*}

When we look at latency between specific node pairs in~\autoref{fig:pairwise_latency_same_subnet_aws}, we can see some fluctuations over 6 hours. While these changes seem to follow the same pattern for AWS, the magnitude of changes is different for distinct node pairs. The GCP data (\autoref{fig:pairwise_latency_same_subnet_gcp}) shows that similar latency shifts can affect some, but not all nodes.

\begin{tcolorbox}[left=1mm,right=1mm,top=1mm,bottom=1mm]
\textbf{Lesson 3:} Substantial, frequent, and rapid round-trip latency fluctuations are common in the cloud. Some latency spikes can reach 2,900$\times$ the average latency and can appear and disappear quickly. 
\end{tcolorbox}

\begin{figure*}[h]
  \centering
  \begin{subfigure}{\smallfigwidth}
    \includegraphics[width=\linewidth]{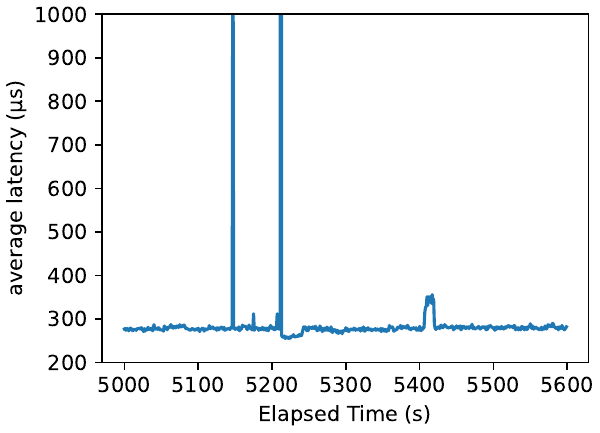}
    \caption{AWS}
    \label{fig:same_subnet_aws_10_min}
  \end{subfigure}
  \begin{subfigure}{\smallfigwidth}
    \includegraphics[width=\linewidth]{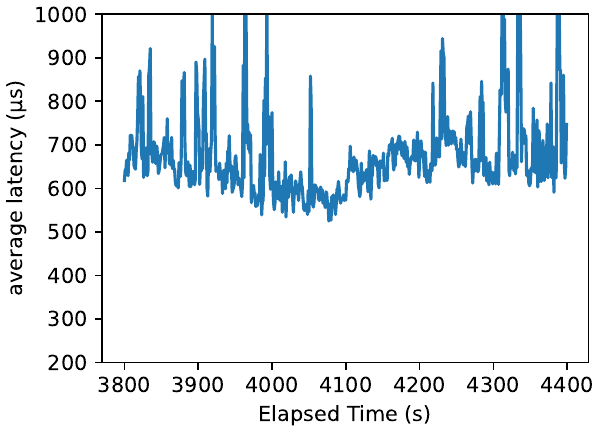}
    \caption{Azure}
    \label{fig:same_subnet_azure_10_min}
  \end{subfigure}  
  \begin{subfigure}{\smallfigwidth}
    \includegraphics[width=\linewidth]{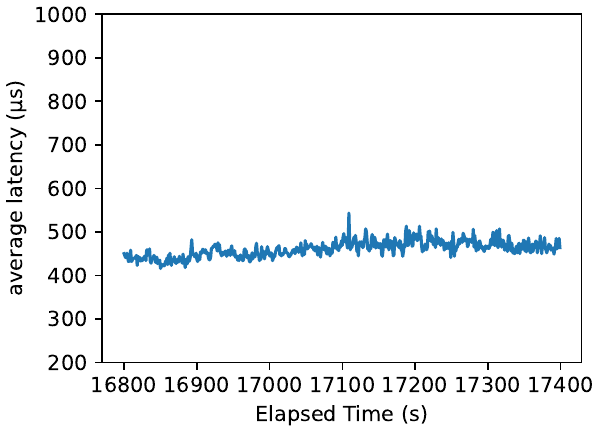}
    \caption{GCP}
    \label{fig:same_subnet_gcp_10_min}
  \end{subfigure}
  
  \caption{Zoom in at some 10-minute interval from~\autoref{fig:latency_same_subnet}. The interval is different for each cloud and shows important latency changes in more detail. (a) focuses on a large latency spike in AWS. (b) shows a latency drop between cyclical features in Azure. (c) shows a 7\% latency increase in GCP.}
  \label{fig:latency_same_subnet_10_min}
  \vspace{-2mm}
\end{figure*}

The Azure observation in ~\autoref{fig:same_subnet_azure} shows several interesting patterns. For one, Azure experiences substantial and frequent latency spikes. Furthermore, Azure latency exhibits noticeable cyclical patterns, as evident in the first hours of the 6-hour run, suggesting a more pronounced instance of noisy-neighbor syndrome during this evaluation run. ~\autoref{fig:same_subnet_azure_10_min} shows one of such cyclical "arches" visible in the first half of ~\autoref{fig:same_subnet_azure} in more detail. \autoref{fig:pairwise_latency_same_subnet_azure} depicts the same subnet/AZ data with round-trip latency between each node pair. We can see the cyclical patterns coming from each node pair (this is especially pronounced for nodes 1.2 and 1.3).

\begin{tcolorbox}[left=1mm,right=1mm,top=1mm,bottom=1mm]
\textbf{Lesson 4:} Cloud may experience cyclical latency fluctuations. 
\end{tcolorbox}

Interestingly enough, the $<$1.2, 1.3$>$ pair seems to show fewer extreme and random fluctuations that show up in pairs $<$1.1, 1.2$>$ and $<$1.1, 1.3$>$, leading to our next lesson:

\begin{tcolorbox}[left=1mm,right=1mm,top=1mm,bottom=1mm]
\textbf{Lesson 5:} Not all nodes of the same type and running identical work behave the same. Communicating with some nodes may innately be more noisy and less predictable than with others. 
\end{tcolorbox}

\begin{figure}
  \centering
  \includegraphics[width=0.85\linewidth]{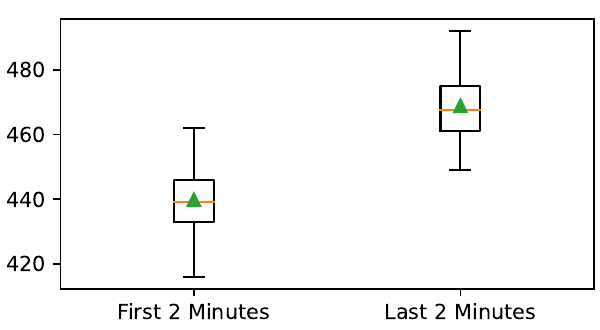}
  \caption{GCP latency for the start and end of ~\autoref{fig:same_subnet_gcp_10_min}. The average latency increases by 7\% in 10 minutes.}
  \label{fig:gcp_two_mins_cmp_boxplot}
  \vspace{-2mm}
\end{figure}

A high-level overview of GCP latency (\autoref{fig:same_subnet_gcp}) shows mostly stable performance with some noticeable spikes throughout the run. More interestingly, GCP round-trip latency in the same subnet of the same AZ seems to sometimes switch between lower and higher latency modes. While this may be difficult to spot in the high-level view~\autoref{fig:same_subnet_gcp_10_min} shows a zoomed-in view on a 10-minute interval with one of such transitions. There, the round-trip latency appears to gradually increase from a median latency of 439 microseconds to 468 microseconds at the end of that 10-minute interval, as shown in~\autoref{fig:gcp_two_mins_cmp_boxplot} that depicts box plots for the first and last two minutes of the 10-minute interval from ~\autoref{fig:same_subnet_gcp_10_min}. The cause of this shift can be seen in~\autoref{fig:pairwise_latency_same_subnet_gcp}, where node pairs $<$1.1, 1.3$>$ and $<$1.2, 1.3$>$ transition from lower round-trip latency to higher latency in a period of 10 minutes starting roughly at the 16,800-second mark. 

\begin{tcolorbox}[left=1mm,right=1mm,top=1mm,bottom=1mm]
\textbf{Lesson 6:} The round-trip latency between any two nodes 
%in the same subnet and AZ 
may change substantially and for the long term in a matter of minutes and without any warning signs. 
\end{tcolorbox}

\begin{figure*}[t]
  \centering
  \begin{subfigure}{\smallfigwidth}
    \includegraphics[width=\linewidth]{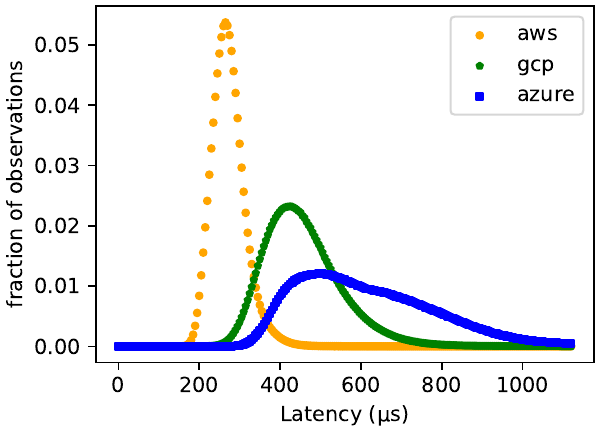}
    \caption{Same Subnet Latency Histogram}
    \label{fig:same_subnet_hist}
  \end{subfigure}
  \begin{subfigure}{\smallfigwidth}
    \includegraphics[width=\linewidth]{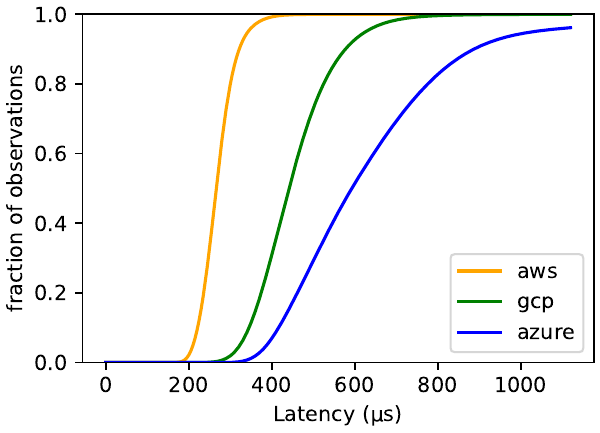}
    \caption{Same Subnet CDF}
    \label{fig:same_subnet_cdf}
  \end{subfigure}
  \begin{subfigure}{\smallfigwidth}
    \includegraphics[width=\linewidth]{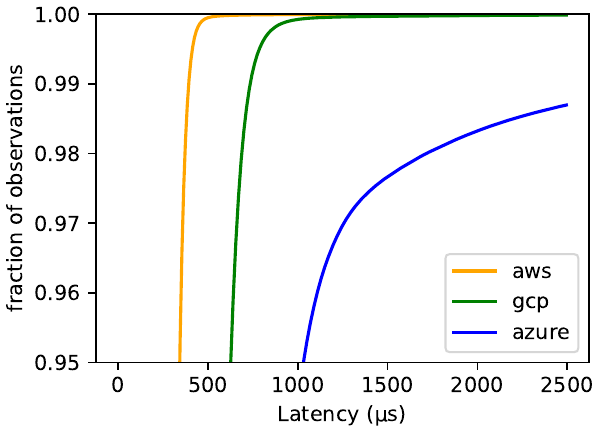}
    \caption{Same Subnet 95\%-100\% CDF}
    \label{fig:same_subnet_cdf_zoom}
  \end{subfigure}
  
  \caption{Same subnet/AZ round-trip latency distribution over a 6-hour interval.}
  \label{fig:latency_distribution_same_subnet}
\end{figure*}

To further study the latency distribution in the same subnet of an AZ, we plot the latency distribution histogram and CDF in ~\autoref{fig:latency_distribution_same_subnet}. The distributions for two clouds, AWS and GCP, appear largely Normal. Azure latency, on the other hand, has a much larger tail toward higher latency. This high tail shows in the latency CDF~\autoref{fig:same_subnet_cdf}. A zoomed version of the CDF, showing the top 5\% of latency observations, indicates that Azure's 99\textsuperscript{th} percentile latency exceeds 2.5 ms. Azure had a 99\textsuperscript{th} percentile of 3.14 ms and 99.99\textsuperscript{th} percentile latency of 21.2 ms. AWS and GCP fare much better, with 99.99\textsuperscript{th} percentile latency of 0.77 ms and 2.79 ms, respectively. 

\begin{tcolorbox}[left=1mm,right=1mm,top=1mm,bottom=1mm]
\textbf{Lesson 7:} Expect high tail latency, with 99\textsuperscript{th} percentile often twice the average latency and 99.99\textsuperscript{th} percentile reaching 25$\times$ the average.
\end{tcolorbox}

%Figure 3: three horizontal panes (a) histogram with aggregated same-subnet data for 3 clouds, (b) histogram for cross-subnet for 3 clouds, (c) histogram for cross AZ for 3 clouds

%~\autoref{fig:latency_agg_hist} shows the latency distribution for each cloud. All of these distributions had long tails, which are cut off for the sake of readability. Of particular interest is that Azure is slightly bimodal when moving across subnets, and both AWS and Azure become strongly bimodal when crossing AZs. 
\subsubsection{Cross-AZ Round-Trip Latency}
\begin{figure*}[t]
  \centering
  \begin{subfigure}{\smallfigwidth}
    \includegraphics[width=\linewidth]{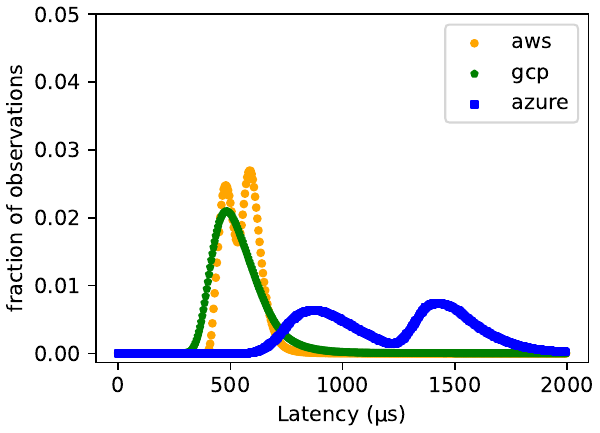}
    \caption{Cross AZ Latency Histogram}
    \label{fig:cross_az_hist}
  \end{subfigure}
  \begin{subfigure}{\smallfigwidth}
    \includegraphics[width=\linewidth]{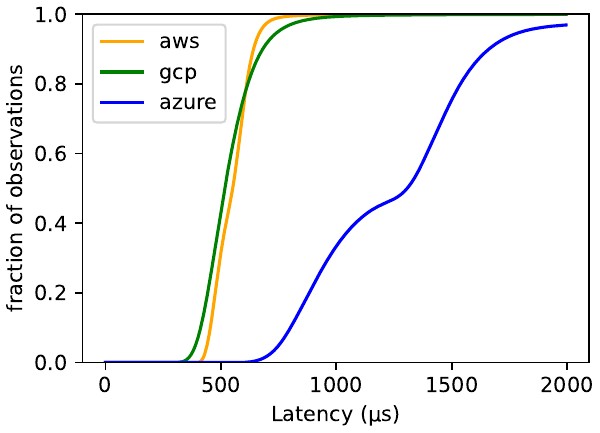}
    \caption{Cross AZ CDF}
    \label{fig:cross_az_cdf}
  \end{subfigure}
  \begin{subfigure}{\smallfigwidth}
    \includegraphics[width=\linewidth]{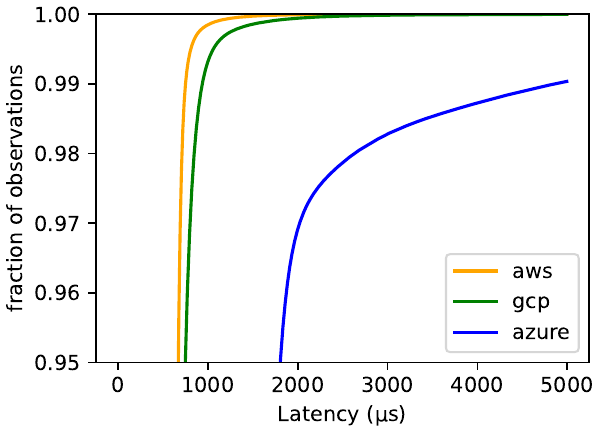}
    \caption{Cross AZ 95\%-100\% CDF}
    \label{fig:cross_subnet_cdf_zoom}
  \end{subfigure}
  
  \caption{Cross-AZ round-trip latency distribution over a 6-hour interval.}
  \label{fig:latency_distribution_cross_az}
\end{figure*}

Many applications and services place their components across different networks and AZs of a region to improve fault tolerance and resilience by ensuring that nodes reside in independent failure domains~\cite{aurora,dynamodb,documentdb}. Such design choices may have an impact on the observed communication latency. The latency distribution for nodes in the same AZ but across different subnets looks nearly indistinguishable from the same-subnet data (\autoref{fig:latency_distribution_same_subnet}), as can be seen in the high-level overview of our same-region data (\autoref{fig:latency_agg_boxplot}). However, the communication latency changes more drastically when placing nodes across different AZs. ~\autoref{fig:latency_distribution_cross_az} shows the round-trip latency distribution for nodes in different AZs over the same 6-hour run as in previous experiments. The most noticeable feature is the bi-modal latency distributions for AWS and Azure. GCP does not experience such bi-modal behavior, but its round-trip latency across AZs and within AZ are much closer together than in other clouds.

The bi-modal nature of AWS and Azure cross-AZ distributions may be the result of different latencies between data centers that support distinct AZs. For example, AZ1 may be closer to AZ2 than AZ3, resulting in slightly lower latency between AZ1 and AZ2 (it takes light 3.3 seconds to travel 1km in a vacuum). Additionally, it is worth noting that for AWS, a single AZ may be supported by multiple data centers~\cite{aws_az_many_dc}, adding a degree of luck to the picture for latency even within the same AZ, as we discuss in~\autoref{sec:vm_placement}. The overview data presented in the~\autoref{fig:latency_agg_boxplot}, the multi-modal nature of cross-AZ latency distributions, and the potential for multi-modal latency for the same AZ due to cross-datacenter communication ~\autoref{sec:vm_placement}, leads us to the following lesson: 

\begin{tcolorbox}[left=1mm,right=1mm,top=1mm,bottom=1mm]
\textbf{Lesson 8:} Communication across AZs has a noticeable penalty compared to communication within the AZ, but the degree of such penalty varies depending on the cloud and factors such as VM placement in the data center or AZ.
\end{tcolorbox}

\subsubsection{Impact of VM Placement}
\label{sec:vm_placement}
~\autoref{fig:pairwise_latency_same_subnet} also shows that different nodes may have substantially different latency throughout their lifetime. This is evident in every cloud we tested. For example, ~\autoref{fig:pairwise_latency_same_subnet_aws} illustrate the AWS with a clear difference in communication latency between node pairs $<1.1, 1.3>$ and $<1.1, 1.2>$. While the difference, on average, does not exceed 20 microseconds, it is apparent during the entire 6-hour run. Moreover, we can observe the latency of all node pairs fluctuating similarly during that experiment. Azure, on ~\autoref{fig:pairwise_latency_same_subnet_azure} has similar behavior, with differences between fast and slow node pairs of over 100 microseconds.  

%~\autoref{fig:three_run_cmp_cross_az} shows a latency distribution for multiple runs conducted at the same time of the data on multiple days. This demonstrates the variability of collected data since 3 runs on the same VMs resulted in noticeably different distributions for all clouds. 

\begin{figure*}[t]
  \centering
  \begin{subfigure}{\smallfigwidth}
    \includegraphics[width=\linewidth]{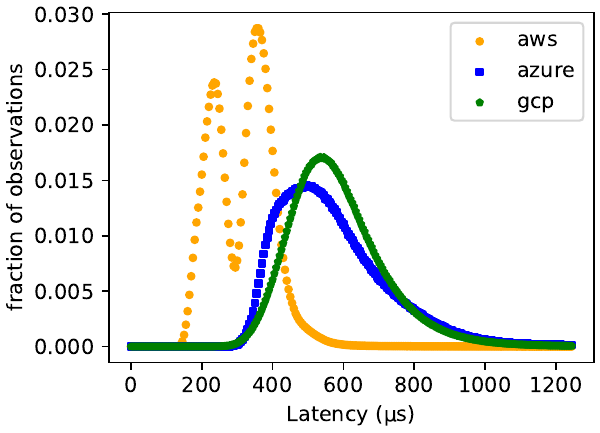}
    \caption{Same Subnet}
    \label{fig:three_run_cmp_same_subnet}
  \end{subfigure}
  \begin{subfigure}{\smallfigwidth}
    \includegraphics[width=\linewidth]{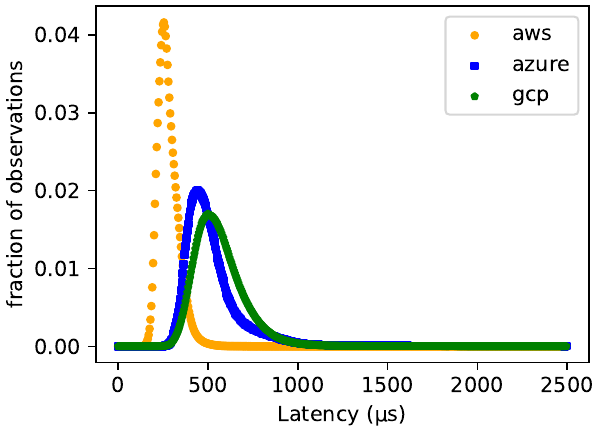}
    \caption{Cross Subnet}
    \label{fig:three_run_cmp_cross_az_azure}
  \end{subfigure}
  \begin{subfigure}{\smallfigwidth}
    \includegraphics[width=\columnwidth]{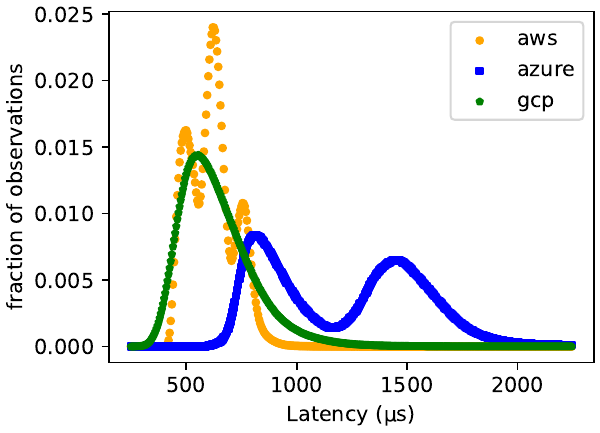}
    \caption{Cross AZ}
    \label{fig:three_run_cmp_cross_az}
  \end{subfigure}
  
  \caption{Comparison for Three 1-hour Runs}
  \label{fig:three_run_cmp}
\end{figure*}

As part of our reproducibility effort, we performed several data collection runs. ~\autoref{fig:three_run_cmp} we aggregate the data from three 1-hour runs conducted on three different weekdays during business hours. We ensure that each run produced the same amount of data to ensure equal weight in the aggregated result. For each run, the VMs were brought up from the shutdown state. We did not use any advanced VM placement configurations besides specifying the AZ to allow the cloud to pick the VM placement within the AZ~\cite{placementAWS,placementAzure,placementGCP}. 

~\autoref{fig:three_run_cmp_same_subnet} shows the aggregated latency distribution across the three 1-hour runs. The most noticeable difference between this experiment and the 6-hour data (\autoref{fig:same_subnet_hist}) is the multi-modal latency in AWS for same-AZ communication. We further inspected this phenomenon, and observed a very pronounced difference between several node pairs, in fact, with different clusters of node pairs corresponding to the peaks in the aggregated histogram data. In our observations, this difference can reach nearly 200 microseconds, a very substantial variation considering the initial run's same AZ average latency of only 260 microseconds. We believe the answer for such variability lies in the AZ structure at AWS. In particular, many distinct data centers may support each AZ~\cite{aws_az_many_dc}. If the allocated VMs in the same AZ happen to reside in different data centers, we expect to incur a latency penalty. Unfortunately, this may present an element of luck for users if they do not resort to tools such as placement groups, representing a serious problem for many academic system evaluations in the cloud, as very little thought is typically given to VM placement in literature. 

\begin{tcolorbox}[left=1mm,right=1mm,top=1mm,bottom=1mm]
\textbf{Lesson 9:} Round-trip communication latency depends on VM placement within the AZ and datacenter, however, tools exist to influence placement\cite{placementAWS,placementAzure,placementGCP}.
\end{tcolorbox}

%Figure 5: Bar graphs for comparing same-subnet setup but with different VM sizes (average + std.error, median, 99th, max)

%Figure 6: Bar graphs for comparing different payload sizes  (average + std.error, median, 99th, max)

\subsubsection{Payload Size}
While our default test used a 1024-byte payload in each message going back and forth, the message size can impact the performance. Bigger payloads may exceed the Network's maximum transmission unit (MTU), requiring the message to split into multiple packets~\cite{aws_mtu}. For all of our clouds, we used the default MTU settings. For AWS nodes in the same regions, the MTU is 1300 bytes~\cite{aws_mtu}, so a 1024-byte payload should fit into a single packet, while larger payloads will be fragmented. Similarly, GCP's MTU between our nodes was 1460 bytes, while Azure's MTU was 1500 bytes.%---the maximum supported value for standard Ethernet v2~\cite{ethv2_mtu}. %We did not use Jumbo Frames. 

%~\autoref{fig:payload_sizes_boxplots} shows the effect of payload size on latency. This is important for systems that can control request size, for instance, due to request batching, since the latency overhead may be important. These figures use data from the "Same Subnet" group to attempt to control for cross-subnet or cross-az latency increases. The experiments were performed one after another on the same VMs, so the results in the figure may be inaccurate due to the long-term latency shifts seen in ~\autoref{fig:pairwise_latency_same_subnet}.

\begin{figure*}[t]
  \centering
  \begin{subfigure}{\smallfigwidth}
    \includegraphics[width=\linewidth]{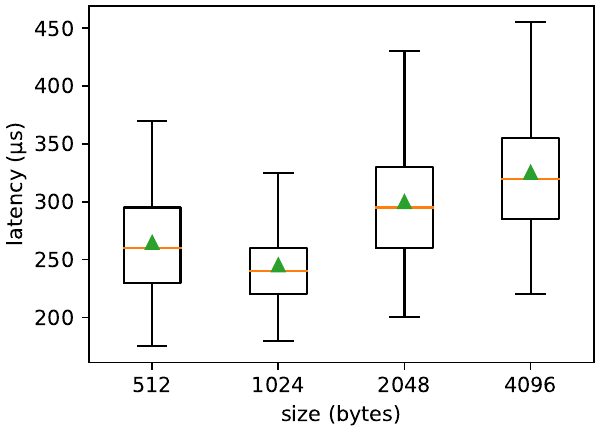}
    \caption{AWS}
    \label{fig:payload_sizes_boxplot_aws}
  \end{subfigure}
  \begin{subfigure}{\smallfigwidth}
    \includegraphics[width=\linewidth]{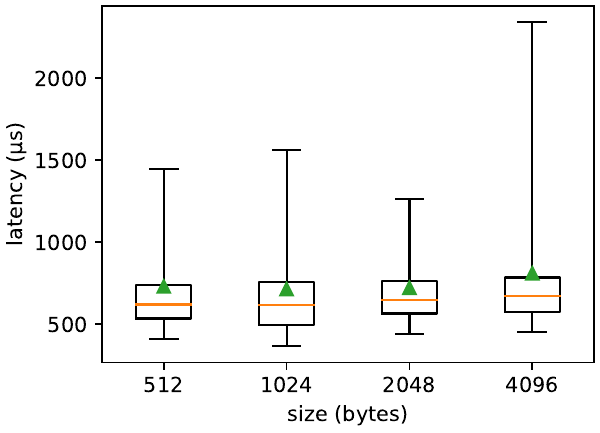}
    \caption{Azure}
    \label{fig:payload_sizes_boxplot_azure}
  \end{subfigure}
  \begin{subfigure}{\smallfigwidth}
    \includegraphics[width=\columnwidth]{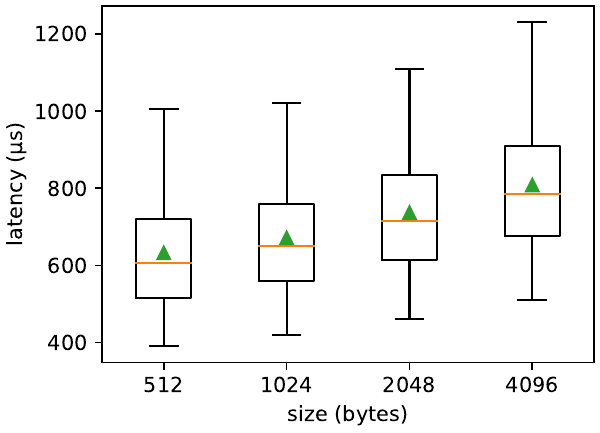}
    \caption{GCP}
    \label{fig:payload_sizes_boxplot_gcp}
  \end{subfigure}
  
  \caption{Round-trip latency over a 1-hour interval for different payload sizes. The data were obtained over 4 hours running back-to-back and are subject to possible daily fluctuations shown in~\autoref{fig:pairwise_latency_same_subnet}.}
  \label{fig:payload_sizes_boxplots}
  \vspace{-2mm}
\end{figure*}

~\autoref{fig:payload_sizes_boxplots} shows the effect of payload size on round-trip latency between VMs in the same subnet/AZ. In this experiment, we ran CLT with each payload size for one hour during business hours on a weekday. We used the same VMs for all payload sizes to avoid the VM-allocation lottery described in~\autoref{sec:vm_placement}. The round-trip latency for GCP (\autoref{fig:payload_sizes_boxplot_gcp}) gradually increases as the payload size grows. Interestingly, AWS shows the lowest average latency for a 1024-byte payload and increased latency for larger messages. The 512-byte message, while comparable to 1024-bytes has more variability. Please note that the difference between 512 bytes and 1024 is small, and falls within the scope of fluctuations we observed over the 6 hours (\autoref{fig:pairwise_latency_same_subnet}). On Azure, we also see a small growth in average and mean latency as the payload increases, although the difference is less pronounced due to higher variation in observed latency. 

\begin{tcolorbox}[left=1mm,right=1mm,top=1mm,bottom=1mm]
\textbf{Lesson 10:} Payload size will likely increase the round-trip latency, especially if the payload exceeds MTU and must be fragmented. While in some cases, the impact on latency may be very pronounced (nearly 200 microseconds median latency difference between 512 and 4096 bytes!), in other cases, it can be masked by daily latency variations. 
\end{tcolorbox} 

\subsubsection{Latency in a Quorum}
Quorum-style communication is a popular method of masking failures and slow nodes in many data-driven systems. In quorum systems, the same operation or request completes multiple times at different nodes for reliability purposes. Typically, one node will send the message/request to its peers, and wait for a sufficient number of replies before considering the operation complete in the quorum. The completion latency, therefore, is determined by the quorum of the fastest nodes. The most popular quorum system is majority quorums, requiring a majority from the set of all nodes to complete each request. The latency of communication rounds in such a quorum system is determined by the fastest majority. For instance, many database systems rely on three-way replication by default~\cite{raft,epaxos,spanner,cockroachdb} and need a quorum of two nodes, allowing the slowest node to still receive and process requests, but masking its ``slowness'' from the users. 

%~\autoref{fig:quorum} shows the impact of quorum sizes on latency. NQ stands for No Quorum and is pairwise communication. Q2/3 means a quorum of 2 out of 3 participating systems. Q3/3 means all 3 participating systems must respond. Q2/3 shows latency benefits because it can accept whichever of its peers is faster. 

\begin{figure*}[t]
  \centering
  \begin{subfigure}{\smallfigwidth}
    \includegraphics[width=\linewidth]{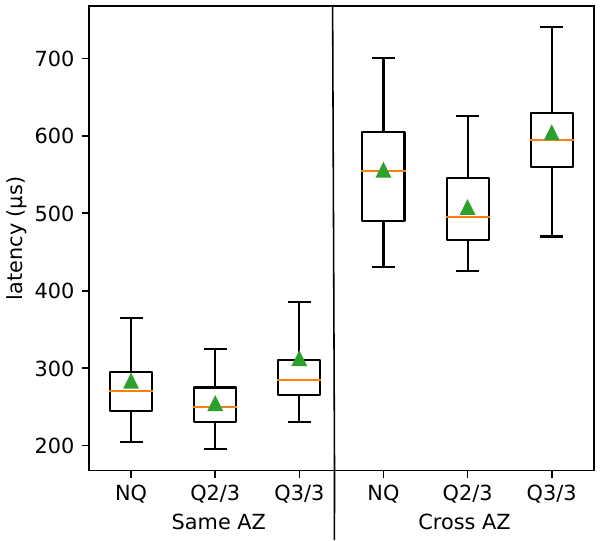}
    \caption{AWS}
    \label{fig:quorum_aws}
  \end{subfigure}
  \begin{subfigure}{\smallfigwidth}
    \includegraphics[width=\linewidth]{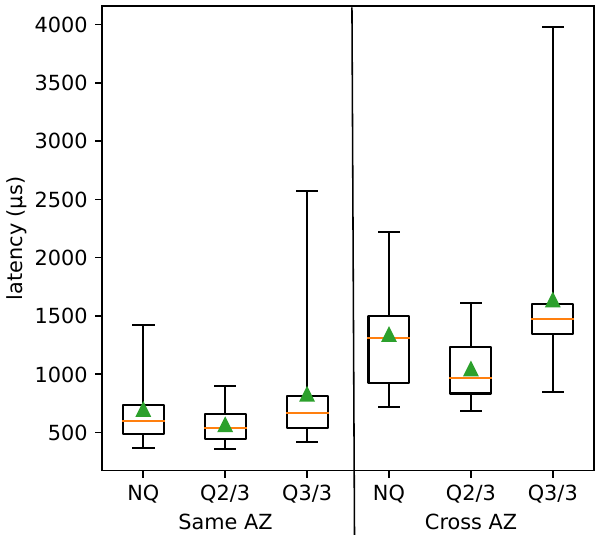}
    \caption{Azure}
    \label{fig:quorum_azure}
  \end{subfigure}
  \begin{subfigure}{\smallfigwidth}
    \includegraphics[width=\linewidth]{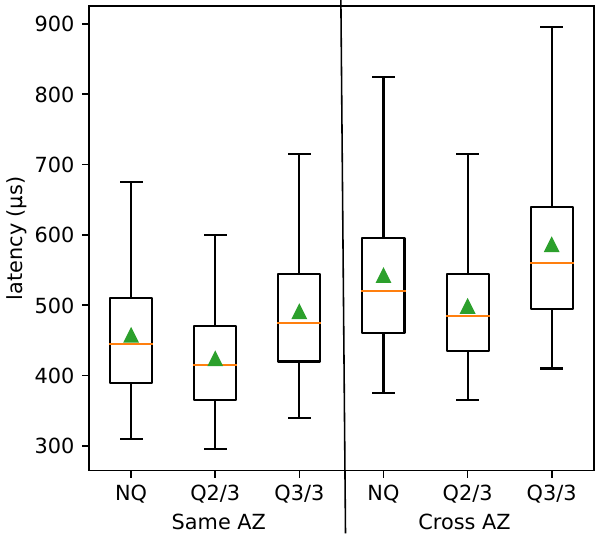}
    \caption{GCP}
    \label{fig:quorum_gcp}
  \end{subfigure}
  
  \caption{Box plots for round-trip latency over a 6-hour interval with quorums. NQ stands for no-quorum distribution, representing a simple node-to-node communication. Q2/3 is a majority quorum of two nodes out of three for each communication round. Q3/3 is an all-quorum, requiring a response from all nodes in each round. }
  \label{fig:quorum}
  \vspace{-2mm}
\end{figure*}

CLT operates in rounds and can measure quorum latency by only looking at the fastest replies in each round. ~\autoref{fig:quorum} shows the impact of quorums on latency. In this figure, we use our default 6-hour dataset and compute the majority quorum (2 out of 3 nodes) round-trip latency, marked as Q-2/3, and all nodes quorum (Q-3/3) for both the same subnet/AZ and cross-AZ communication.

The majority quorum can filter the slowest node in each round, resulting in better overall latency than waiting for all nodes in a round (i.e., a quorum of all nodes, Q-3/3). The majority quorum also has better latency distribution than simple round-trip node-to-node communication---the node-to-node latency spikes impact the distribution and drive the average higher, while quorums ``forget'' the slowest node in each round, lowering the average. Such masking effect is especially noticeable at extreme tail latency: the 99.999\textsuperscript{th} percentile latency for quorums at AWS was 0.85ms in the same AZ, while the same percentile latency for individual node-to-node observations was 490.72 ms, a 577$\times$ difference. Similar improvements occur across all cloud providers and AZs. 

\subsection{Cross-region Communication}
\begin{figure}[t]
  \centering
  \begin{subfigure}{\linewidth}
    \includegraphics[width=\linewidth]{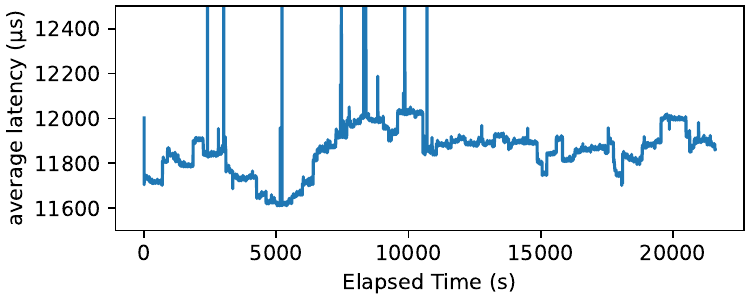}
    \caption{AWS}
    \label{fig:cross_region_east_east_aws}
  \end{subfigure}
  
  \begin{subfigure}{\linewidth}
    \includegraphics[width=\linewidth]{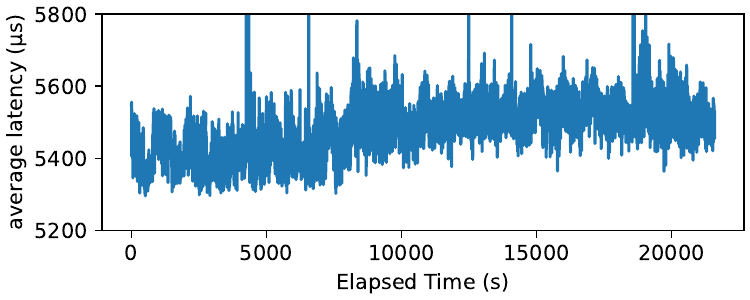}
    \caption{Azure}
    \label{fig:cross_region_east_east_azure}
  \end{subfigure}
  
  \begin{subfigure}{\linewidth}
    \includegraphics[width=\linewidth]{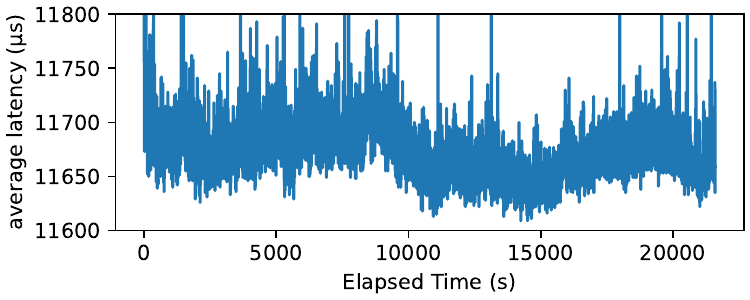}
    \caption{GCP}
    \label{fig:cross_region_east_east_gcp}
  \end{subfigure}
  
  \caption{Cross-region East US to East US (East1---East2) round-trip latency over the 6-hour interval.}
  \label{fig:cross_region_east_east}
    \vspace{-2mm}
\end{figure}

%Figure 1: three vertical panes, one for each cloud, showing a birds-eye view (i.e., 15 or 30-second granularity) of the latency in the following pairs: (1.x <-> 2.1, 1.x <-> 3.1, 2.1 <-> 3.1)
Our 6-hour dataset includes nodes at 3 different regions, allowing us to examine the variability of cross-region communication. ~\autoref{fig:cross_region_east_east} shows the cross-region latency between two regions in the eastern United States (East1 and East2). These regions are relatively close to each other and have smaller latencies. Please note that due to the different geographical placement, the latencies between the two East regions in different clouds are very distinct. Instead of comparing latencies, we focus on the stability of the cross-region latency. The AWS cross-region latency, shown in~\autoref{fig:cross_region_east_east_aws} appears ``jagged'' with sudden and abrupt shifts up and down. Excluding a few occasional spikes, the difference between the lowest and highest sufficiently stable and prolonged periods of operation is around 1.5 ms or roughly 10\% of the average latency between the two regions. Both Azure and GCP do not have the ``jagged'' latency and instead show frequent fluctuations up and down with some longer-term trends.

%~\autoref{fig:cross_region_east_east} shows the latency of cross-region communication between each of the clouds for two regions in the eastern United States. The pairs of regions are not geographically located the same distance apart for each cloud, so they should not be directly compared. The scale on each sub-figure is different to allow more detail to be visible. This allows the "step" pattern on ~\autoref{fig:cross_region_east_east_aws} to be visible. 

%~\autoref{fig:cross_region_east_west} shows the latency of cross-region communication between each of the clouds for two regions, one in the eastern United States, and one in the western United States. The pairs of regions are not geographically located the same distance apart for each cloud, so they should not be directly compared. The scale on each sub-figure is different to allow more detail to be visible. This allows the "step" pattern that previously appeared only in ~\autoref{fig:cross_region_east_east_aws} to be visible for all clouds. 

\begin{figure}[t]
  \centering
  \begin{subfigure}{\linewidth}
    \includegraphics[width=\linewidth]{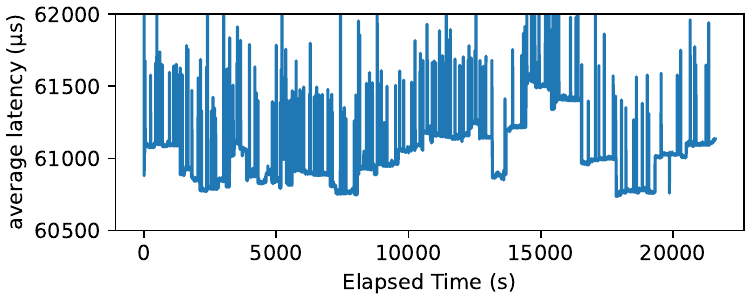}
    \caption{AWS}
    \label{fig:cross_region_east_west_aws}
  \end{subfigure}
  
  \begin{subfigure}{\linewidth}
    \includegraphics[width=\linewidth]{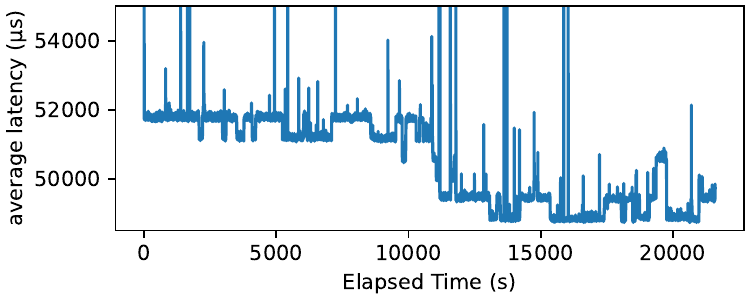}
    \caption{Azure}
    \label{fig:cross_region_east_west_azure}
  \end{subfigure}
  
  \begin{subfigure}{\linewidth}
    \includegraphics[width=\linewidth]{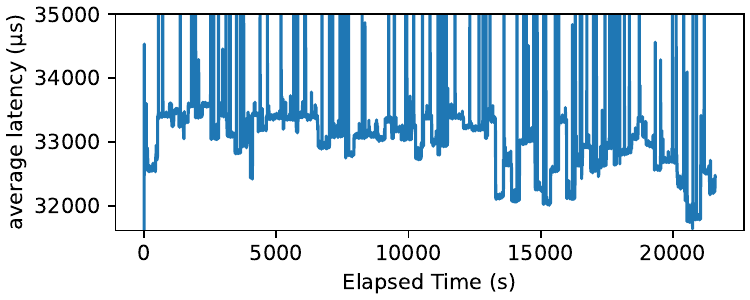}
    \caption{GCP}
    \label{fig:cross_region_east_west_gcp}
  \end{subfigure}
  
  \caption{Cross-region East US to West US round-trip latency over the 6-hour interval.}
  \label{fig:cross_region_east_west}
  \vspace{-2mm}
\end{figure}

Looking at larger distances between regions, ~\autoref{fig:cross_region_east_west} illustrates the latency between the East (East1) and West regions of our cloud deployments. In this case, all three clouds exhibit the ``jagged'' step-like latency pattern with abrupt shifts up or down. Similarly, there is a noticeable difference between high- and low-latency modes. For instance, in Azure, it is around 3ms, as latency largely shifted from 52ms to 49 ms, while AWS shows a difference of roughly 0.75 ms, excluding spikes.

The jagged patterns for cross-region communication may indicate the different routes packets predominantly take at different times depending on load and congestion in the cloud provider's cross-region network. Nevertheless, similar to latency within the region and within the same AZ, tenants need to expect substantial variation. During one of our reproducibility runs, we encountered a severe issue on the Azure network, impacting a single East1 node's ability to communicate with the West region. In that run, the round-trip latency for communication between nodes 1.1 and 3.1 went above 5 seconds(!) and remained high (above 4 seconds round-trip) for several minutes. At the same time, no other nodes in the East1 region were affected; node 1.1 continued normal message exchange with all peers except 3.1, while other nodes could communicate with node 3.1 without apparent problems. 

\begin{tcolorbox}[left=1mm,right=1mm,top=1mm,bottom=1mm]
\textbf{Lesson 11:} Cross-regional communication often exhibits a ``jagged'' pattern with abrupt changes up and down that can nearly instantaneously change average round-trip latency by as much as a few milliseconds.
\end{tcolorbox} 

%Related
\section{Related Work}
\label{sec:related}
\textbf{Network Simulators.} Simulation is a popular methodology for analyzing network performance~\cite{sim1}. However, it does not necessarily apply as meaningfully when trying to replicate the behavior in the cloud. Various tools such as the ns-3 simulator\cite{ns3}, OMNeT++\cite{omnet}, and NetSim\cite{netsim} offer the ability to simulate reproducible results for a network system, given the network topology and some information about the system, such as network nodes, channels, protocols, and traffic (data flow)~\cite{netsim2}. These tools allow engineers who know the infrastructure of existing systems to model the behavior of communication and analyze the results for performance metrics. However, such simulations have limitations compared to real-time observations. 

First, the simulations rely on knowledge of the network topology and specification. For a system that the user has full knowledge about and control in implementation, this is not a problem. However, this is not likely in the public cloud, as users are not aware of the network and the exact placement of nodes in the system. Therefore, these types of simulations can only provide realistic estimates within a margin and are less accurate depending on the amount of information known about how the system is structured. In contrast to this, we perform real-world measurements in the cloud, and while our data has limitations (\autoref{sec:validity}), it provides a reasonable snapshot of cloud behavior. 

Second, the simulations assume a controlled environment. In contrast, cloud systems operate as a global hive mind in which thousands of users contribute to traffic in ways that we have limited control over. We can make predictions about usage, which we can include in simulations, and we can mitigate these hypothesized impacts to the best of our abilities, but ultimately we can only use our best estimations to simulate "real" or "average" cloud usage. Even then, creating a simulation that accurately replicates the amount of traffic that cloud systems handle is expensive in terms of resources and processing time, making it impractical for analyzing longer periods.

\textbf{Similar Systems.} Communication latency is an important metric concerning networks and distributed systems that rely on them\cite{latency_cloud_impact}. As a result, plenty of tools exist for the sole purpose of pinging various cloud service machines across different regions and countries and recording the latency~\cite{websites_ping1, websites_ping2, websites_ping3}. These tools are helpful for testing the response of these services at a given moment, typically focusing on pinging anywhere from five to over 40 different machines per service across the world from a single vantage point, often the user's location. Rather than measuring communication from a static point or end-user to a given region, we focus on observing the latency variability between nodes within the cloud. 

A few studies look at the latency in the cloud. For instance, Tomanek et al., ~\cite{latency_measure_1} examines the latency observed from multiple outside vantage points. That study also focuses on the temporal component, comparing the data from 2013 and 2016. In contrast, our work looks at fine-grained latency changes and predictability within the cloud. Other systems looked at some application-specific aspects of latency in the cloud, such as gaming~\cite{cloud_gaming}. Some studies take a more holistic approach and compare cloud providers from multiple relevant performance metrics, including latency~\cite{latency_measurement2}. Others focus on only the network component of end-to-end latency by measuring the delays on relevant packets at the smart switch~\cite{in_network_rtt}. The reliance on smart switches also makes such an approach infeasible for cloud deployment by tenants. Other work, such as Jain et al., ~\cite{jainSkyplaneOptimizingTransfer2023} focus on bandwidth. In this work we assume sufficient bandwidth because cloud providers typically provide a way to purchase more bandwidth if it becomes necessary, making bandwidth a matter of how much money one is willing to spend ~\cite{aws_ec2_bandwidth, azure_instance_bandwidth, gcp_instance_bandwidth}.

%Conclusion
\section{Conclusion}
\label{sec:conclusion}
Communication timing plays a critical role in many systems, with many systems using timing assumptions for operation- or performance-critical tasks. However, these timing assumptions are often estimated without a proper understanding of the environment in which systems may be deployed. In this paper, we design and develop a simple tool to study communication latency between nodes in the cloud environment. With our tool, we showed the fickle nature of the cloud- - latency between nodes or VMs can change abruptly and without warning. These changes can be very substantial, as much as thousands of times the average latency. However, they can also be less dramatic (i.e., 20\% increase or decrease) but long-term, lasting for hours or more. While we cannot possibly study all imaginable deployment scenarios for systems in the cloud, we believe our work provides a substantial introduction to the variability of cloud communication. 

Such communication latency variability may impact systems and applications running in the cloud, especially the ones designed for low latency. Such systems often have knobs and settings controlling the expected communications delays and variations. Our work sets the baseline for such timing parameters and presents a step towards empirically setting them for the most efficient operation in a particular cloud or network. 

The source code is available at 
\href{https://github.com/UNH-DistSyS/UNH-CLT}{https://github.com/UNH-DistSyS/UNH-CLT}.

\bibliographystyle{plain}
\bibliography{acharapko,noah,marielle,owen,links}

%Appendix
\newpage
\section{Appendix}

This appendix contains supplemental figures which we thought were interesting or potentially helpful, but were not important enough for the main body of work. 

%3 runs comparison 1: cross subnet
% \begin{figure*}[t]
%   \centering
%   \begin{subfigure}{\smallfigwidth}
%     \includegraphics[width=\linewidth]{figures/processed_latency_figures/three_run_cmp_cross_subnet_aws.pdf}
%     \caption{AWS}
%     \label{fig:three_run_cmp_cross_subnet_aws}
%   \end{subfigure}
%   \begin{subfigure}{\smallfigwidth}
%     \includegraphics[width=\linewidth]{figures/processed_latency_figures/three_run_cmp_cross_subnet_azure.pdf}
%     \caption{Azure}
%     \label{fig:three_run_cmp_cross_subnet_azure}
%   \end{subfigure}
%   \begin{subfigure}{\smallfigwidth}
%     \includegraphics[width=\columnwidth]{figures/processed_latency_figures/three_run_cmp_cross_subnet_gcp.pdf}
%     \caption{GCP}
%     \label{fig:three_run_cmp_cross_subnet_gcp}
%   \end{subfigure}
  
%   \caption{Cross Subnet Comparison for Three 1-hour Runs}
%   \label{fig:three_run_cmp_cross_subnet}
% \end{figure*}

\begin{figure*}[h]
  \centering
 \begin{subfigure}{\textwidth}
    \centering
    \includegraphics[width=\linewidth,height=0.25\paperheight,keepaspectratio]{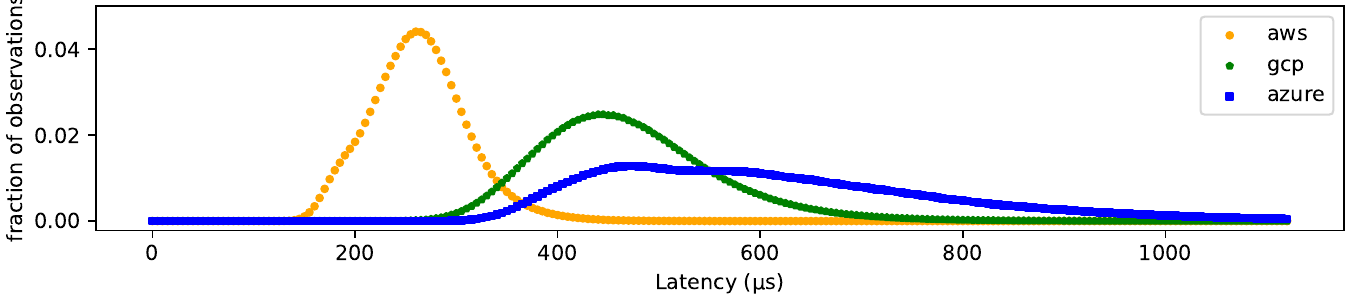}
    \caption{Cross Subnet Latency Histogram}
    \label{fig:cross_subnet_hist}
  \end{subfigure}
  \begin{subfigure}{\textwidth}
    \centering
    \includegraphics[width=\linewidth,height=0.25\paperheight,keepaspectratio]{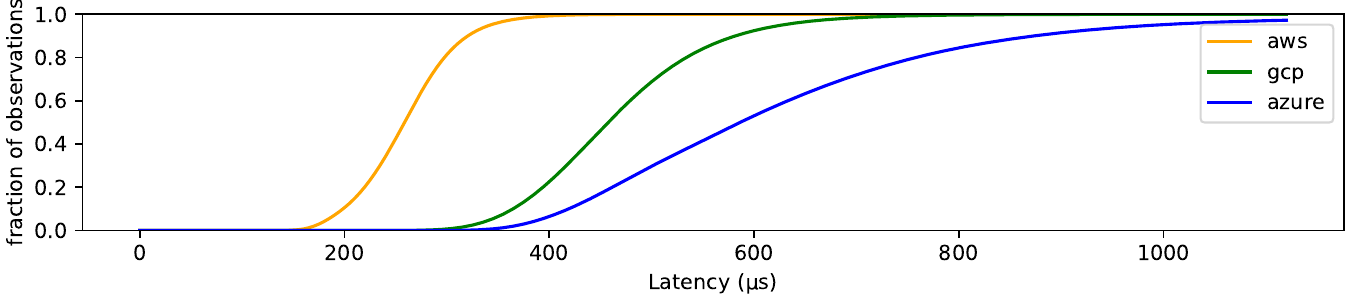}
    \caption{Cross Subnet CDF}
    \label{fig:cross_subnet_cdf}
 \end{subfigure}
  \begin{subfigure}{\textwidth}
    \centering
    \includegraphics[width=\linewidth,height=0.25\paperheight,keepaspectratio]{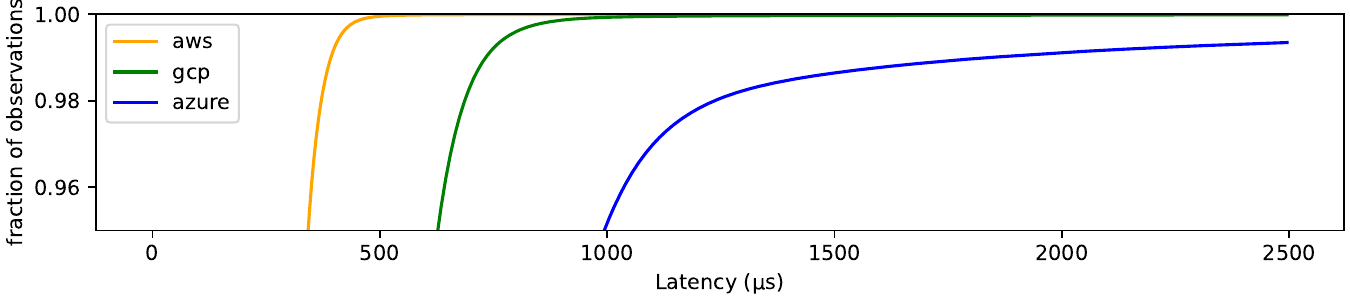}
    \caption{Cross Subnet 95\%-100\% CDF}
    \label{fig:cross_subnet_cdf_zoom}
  \end{subfigure}
 
  \caption{Cross-subnet round-trip latency over the 6-hour interval.}
  \label{fig:cross_subnet}
\end{figure*}

~\autoref{fig:cross_subnet} shows various aggregate measures of latency over the six hours. These are grouped by cloud, but we would like to reinforce that comparing between clouds as far as latency numbers are concerned is specific to our particular run, using the same regions and VM sizes, at a given time. However, ~\autoref{fig:cross_subnet_hist} shows some interesting behavior, as Azure is showing a slightly bimodal distribution, whereas AWS and GCP are much more normally distributed. 

\begin{figure*}[h]
  \centering
 \begin{subfigure}{\textwidth}
    \centering
    \includegraphics[width=\linewidth,height=0.25\paperheight,keepaspectratio]{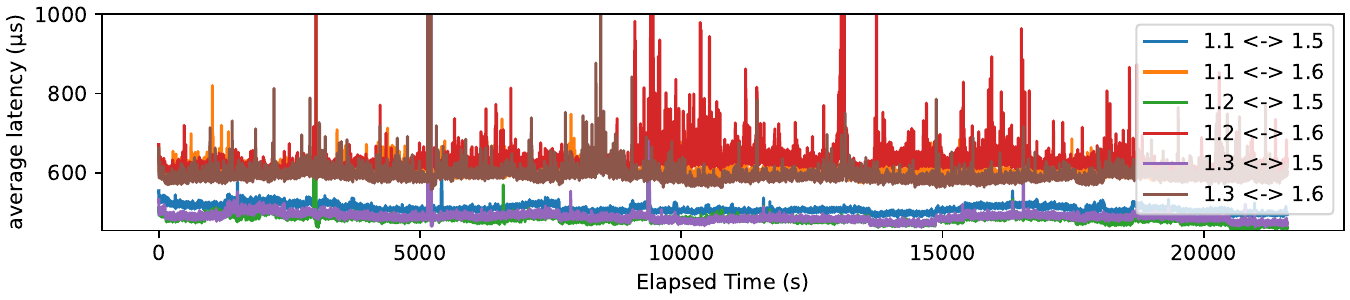}
    \caption{AWS}
    \label{fig:pairwise_cross_az_aws}
  \end{subfigure}
  \begin{subfigure}{\textwidth}
    \centering
    \includegraphics[width=\linewidth,height=0.25\paperheight,keepaspectratio]{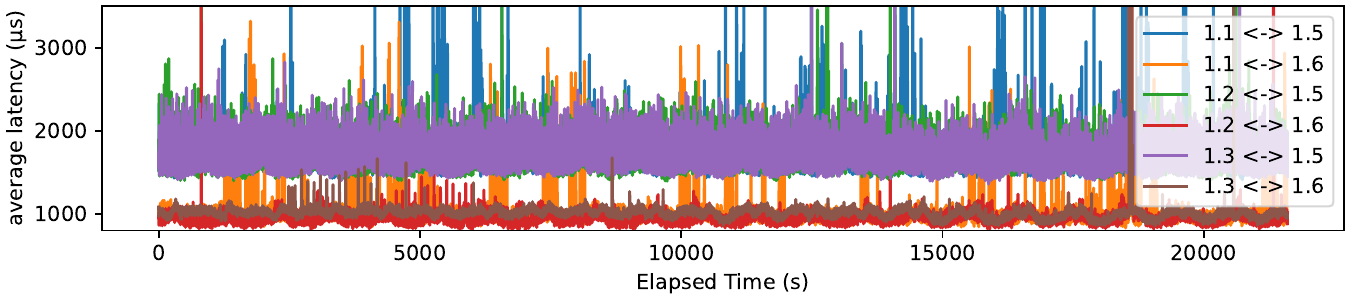}
    \caption{Azure}
    \label{fig:pairwise_cross_az_azure}
 \end{subfigure}
  \begin{subfigure}{\textwidth}
    \centering
    \includegraphics[width=\linewidth,height=0.25\paperheight,keepaspectratio]{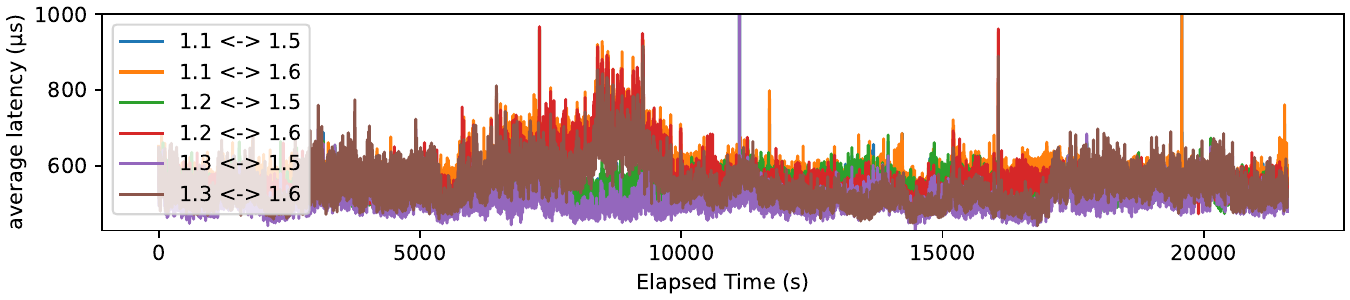}
    \caption{GCP}
    \label{fig:pairwise_cross_az_gcp}
  \end{subfigure}
 
  \caption{Pairwise Cross-AZ round-trip latency over the 6-hour interval.}
  \label{fig:pairwise_cross_az}
\end{figure*}

~\autoref{fig:pairwise_cross_az_aws} displays clear stratification, with all pairs involving 1.5 having noticeably lower latency than those involving 1.6. Since this is cross-AZ communication we hypothesize that this results from the network topology and the physical distance between the nodes. ~\autoref{fig:pairwise_cross_az_azure} exhibits similar behavior to ~\autoref{fig:pairwise_cross_az_gcp}, except pairs with 1.6 are lower latency than pairs with 1.5. ~\autoref{fig:pairwise_cross_az_gcp} is less clear visually, but there is a 40 $\mu$s  difference in mean latency.

\begin{tcolorbox}[left=1mm,right=1mm,top=1mm,bottom=1mm]
\textbf{Lesson 12:} Different pairs AZs in the same region may have different latencies between them. It may be worthwhile to try several combinations of AZs to find the one with the lowest latency for your usecase.
\end{tcolorbox}

\begin{figure*}[h]
  \centering
 \begin{subfigure}{\textwidth}
    \centering
    \includegraphics[width=\linewidth,height=0.25\paperheight,keepaspectratio]{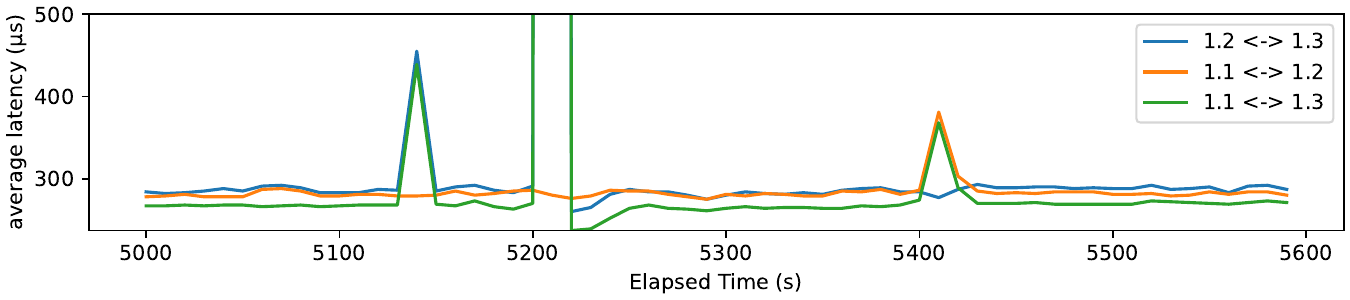}
    \caption{AWS}
    \label{fig:pairwise_same_subnet_zoom_aws}
  \end{subfigure}
  \begin{subfigure}{\textwidth}
    \centering
    \includegraphics[width=\linewidth,height=0.25\paperheight,keepaspectratio]{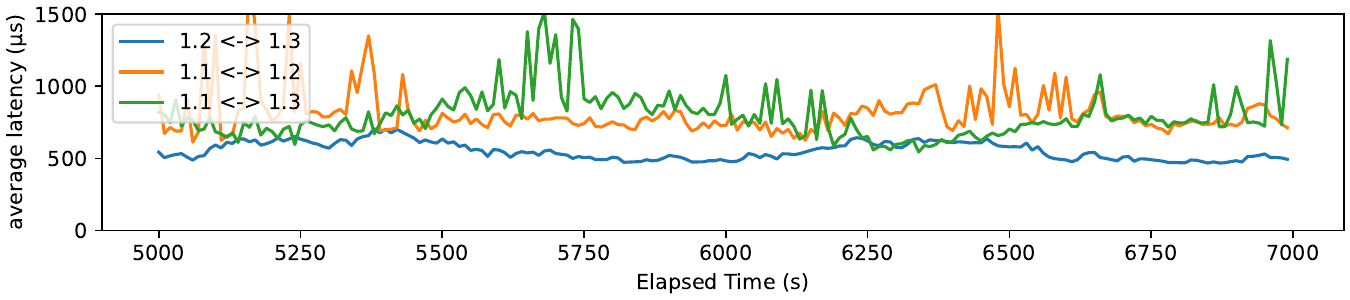}
    \caption{Azure}
    \label{fig:pairwise_same_subnet_zoom_azure}
 \end{subfigure}
  \begin{subfigure}{\textwidth}
    \centering    \includegraphics[width=\linewidth,height=0.25\paperheight,keepaspectratio]{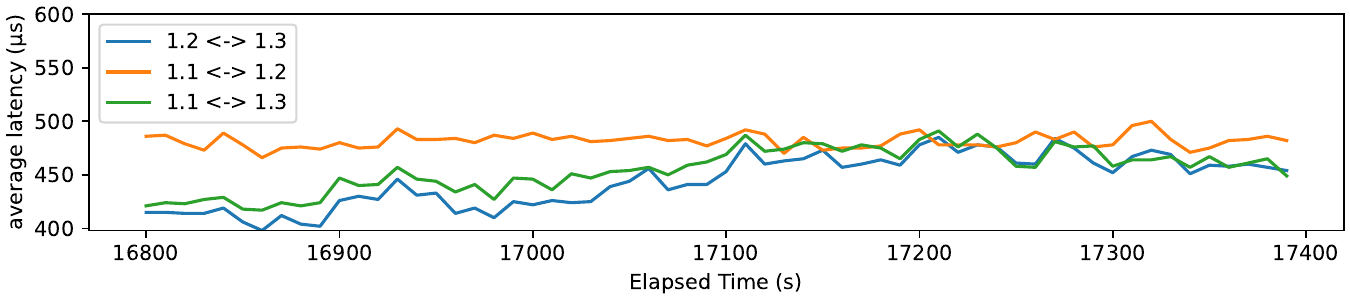}
    \caption{GCP}
    \label{fig:pairwise_same_subnet_zoom_gcp}
  \end{subfigure}
 
  \caption{Pairwise Same Subnet round-trip latency over the 10-minute interval.}
  \label{fig:pairwise_same_subnet_zoom}
\end{figure*}

~\autoref{fig:pairwise_same_subnet_zoom_aws} depicts the large latency spike at 5200 seconds for AWS Same Subnet, also depicted in \autoref{fig:same_subnet_aws_10_min}, showing that 1.1 to 1.2 was unaffected and that 1.3 was likely the culprit. ~\autoref{fig:pairwise_same_subnet_zoom_azure} shows part of one of the "arches" that are present in ~\autoref{fig:same_subnet_azure}, showing that the slight increase in 1.2 to 1.3 combined with the increasing 1.1 to 1.3 likely led to the observed "arch" shape for this area. ~\autoref{fig:pairwise_same_subnet_zoom_gcp} shows that the gradual increase depicted in ~\autoref{fig:same_subnet_gcp_10_min} was the result of the pairs involving 1.3 increasing in latency. As an aside, we also determined that the GCP AZ containing 1.5 was at absolute most 43 miles away from the primary AZ assuming instant network equipment processing, no OS network stack latency and the speed of light in fiber-optics being the same as in a vacuum, so it is likely much closer.  

\newcolumntype{a}{>{\columncolor{lightgray}}{l}}

\begin{table*}
\footnotesize
\centering
\begin{tabular}{|a|r|r|r|r|r|r|r|r|r|}
\hline
\rowcolor{lightgray}
Cloud & \multicolumn{3}{|l|}{AWS} & \multicolumn{3}{|l|}{Azure} & \multicolumn{3}{|l|}{GCP} \\ \hline
\rowcolor{lightgray}
& Same Subnet & Cross Subnet & Cross Az & Same Subnet & Cross Subnet & Cross Az & Same Subnet & Cross Subnet & Cross Az \\ \hline
median ($\mu$s) &       270.0 &        260.0 &    555.0 &       595.0 &        590.0 &   1310.0 &       445.0 &        460.0 &    520.0 \\ \hline
p5 ($\mu$s)    &       215.0 &        185.0 &    445.0 &       390.0 &        395.0 &    755.0 &       325.0 &        345.0 &    395.0 \\ \hline
p25 ($\mu$s)   &       245.0 &        230.0 &    490.0 &       485.0 &        485.0 &    925.0 &       390.0 &        410.0 &    460.0 \\ \hline
mean ($\mu$s)   &       283.0 &        279.0 &    555.4 &       696.7 &        661.9 &   1339.5 &       457.2 &        480.1 &    542.0 \\ \hline
p90 ($\mu$s)   &       325.0 &        325.0 &    645.0 &       890.0 &        875.0 &   1665.0 &       580.0 &        585.0 &    685.0 \\ \hline
p95 ($\mu$s)   &       345.0 &        345.0 &    670.0 &      1035.0 &        995.0 &   1810.0 &       630.0 &        630.0 &    750.0 \\ \hline
p99 ($\mu$s)   &       395.0 &        395.0 &    755.0 &      3135.0 &       1845.0 &   4865.0 &       740.0 &        735.0 &    940.0 \\ \hline
p999 ($\mu$s)  &       470.0 &        470.0 &   1105.0 &      9795.0 &       8925.0 &  16670.0 &       955.0 &        945.0 &   1690.0 \\ \hline
p9999 ($\mu$s) &       760.0 &        745.0 &   1905.0 &     21200.0 &      21705.0 &  23085.0 &      2780.0 &       3210.0 &   3390.0 \\ \hline
p99999 ($\mu$s) &    486136.2 &     619828.0 &  11080.3 &     26797.0 &      35115.0 &  50710.0 &      6227.0 &     390171.4 &  10375.0 \\ \hline
max ($\mu$s)  &    832830.0 &     849470.0 & 832830.0 &    119145.0 &     125500.0 & 137960.0 &     28335.0 &     776095.0 & 328685.0 \\ \hline
\end{tabular}
\centering
\caption{Latency statistics by cloud and network group.}
\label{fig:table_latency_statistics}
\end{table*}

\end{document}